\newif\ifaastex\aastexfalse

\ifaastex
\documentclass[preprint]{aastex}

\else
\documentclass[apj,iop]{emulateapj}
\renewcommand{\epsscale}[1]{}

\fi

\usepackage{graphicx}
\usepackage{amssymb}

\makeatletter

\usepackage{natbib}

\newcommand{\htwo}{H$_2$}

\newcommand{\icfourfourthree}{IC~443}

\newcommand{\vlsr}{\ensuremath{v_{\mathrm{LSR}}}}
\newcommand{\kms}{\ensuremath{\mathrm{km\,s}^{-1}}}

\newcommand{\hcop}{{HCO}$^{+}$}
\newcommand{\hcn}{{HCN}}

\newcommand{\cmthree}{\ensuremath{\rm{cm}^{-3}}}

\newcommand{\msun}{\ensuremath{{M}_{\odot}}}

\newcommand{\hms}[3]{#1$^{\mathrm h}$#2$^{\mathrm m}$#3$^{\mathrm s}$}
\newcommand{\hmss}[4]{#1$^{\mathrm h}$#2$^{\mathrm m}$#3$\fs$#4}

\newcommand{\dms}[3]{#1$\arcdeg$#2$\arcmin$#3$\arcsec$}

\newcommand{\twelveco}{$^{12}${CO}}

\newcommand{\minusthreecloud}{$-3\ \kms$ clouds}
\newcommand{\plussixcloud}{$+5\ \kms$ clouds}

\newcommand{\paperone}{LEE08}

\newcommand{\SIC}{SC}
\newcommand{\SICs}{SCs}

\shorttitle{Identification of Ambient Molecular Clouds around IC~443}
\shortauthors{Lee, J.-J. et al.}

\makeatother
\makeatletter

\begin{document}

\bibliographystyle{apj}

\title{IDENTIFICATION OF AMBIENT MOLECULAR CLOUDS ASSOCIATED WITH GALACTIC
  SUPERNOVA REMNANT IC~443}

\author{Jae-Joon Lee,\altaffilmark{1,2},
Bon-Chul Koo\altaffilmark{3},
Ronald L. Snell\altaffilmark{4},
Min S. Yun\altaffilmark{4},
Mark H. Heyer\altaffilmark{4},
and
Michael G. Burton\altaffilmark{5}
}

\altaffiltext{1}{Korea Astronomy and Space Science Institute, Daejeon, Republic of Korea 305-348}
\altaffiltext{2}{leejjoon@kasi.re.kr}
\altaffiltext{3}{School of Physics and Astronomy, FPRD, Seoul National
  University, Seoul 151-742, Korea}
\altaffiltext{4}{University of Massachusetts, Department of Astronomy,
  Amherst, MA 01003, USA}
\altaffiltext{5}{School of Physics, University of New South Wales,
  Sydney, NSW 2052, Australia}

\begin{abstract}
  The Galactic supernova remnant (SNR) \icfourfourthree\ is one of the
  most studied core-collapse SNRs for its interaction with molecular
  clouds.
However, the ambient
molecular clouds with which  \icfourfourthree\ is interacting have
not been thoroughly studied and remain poorly understood.
  Using Five College Radio Astronomy Observatory 14m telescope, we
  obtained fully sampled maps of $\sim 1\arcdeg \times 1\arcdeg$
  region toward \icfourfourthree\ in the \twelveco\ $J=1-0$ and \hcop\ $J=1-0$
  lines.
  In addition to the previously known molecular
  clouds in the velocity range
  $\vlsr = -6$ to $-1\ \kms$ (\minusthreecloud), our observations
  reveal two new ambient molecular cloud components: small ($\sim
  1\arcmin$) bright clouds in $\vlsr = -8$ to $-3\ \kms$ (\SICs), and
  diffuse clouds in $\vlsr = +3$ to $+10\ \kms$
  (\plussixcloud).
  Our data also reveal the detailed kinematics of the shocked
  molecular gas in \icfourfourthree, however the focus of this paper is
  the physical relationship between the shocked clumps and
  the ambient cloud components.
  We find strong
  evidence that the \SICs\ are associated with the shocked
  clumps. This is supported by the positional coincidence of the
  \SICs\ with shocked clumps and other tracers of shocks.
  Furthermore, the kinematic features of some shocked clumps suggest
  that these are the ablated material from the \SICs\ upon the
  impact of the SNR shock.
  The \SICs\ are interpreted as dense cores of parental molecular
  clouds that survived the destruction by the pre-supernova evolution
  of the progenitor star or its nearby stars.  We propose that the
  expanding SNR shock is now impacting some of the remaining cores and
  the gas is being ablated and accelerated producing the shocked
  molecular gas.
  The morphology of the \plussixcloud\ suggests an association with
  \icfourfourthree. On the other hand, the \minusthreecloud\
  show no evidence for interaction.

\end{abstract}


\section{INTRODUCTION}

\label{intro}

Massive, newborn stars subject the surrounding parent molecular cloud
to disruptive ultraviolet radiation and stellar winds
\citep[e.g.,][]{1999ApJ...511..798C}.  For
O-type stars, a large area around them is expected to be
cleared of molecular clouds, thus when the stars explode
as supernovae (SNe), their remnants will spend a significant
 time expanding inside the hot bubble created by the
progenitors.
On the other hand,
the destruction of natal molecular clouds would not be as effective for
early B stars, and the supernova remnants (SNRs) of these less massive
stars could be able to interact with the molecular clouds that survived the
destruction.  Therefore, the environment of core-collapse SNRs not
only affects the evolution of SNRs, but also provides information
about their progenitor star.

\begin{figure*}[t]
\plotone{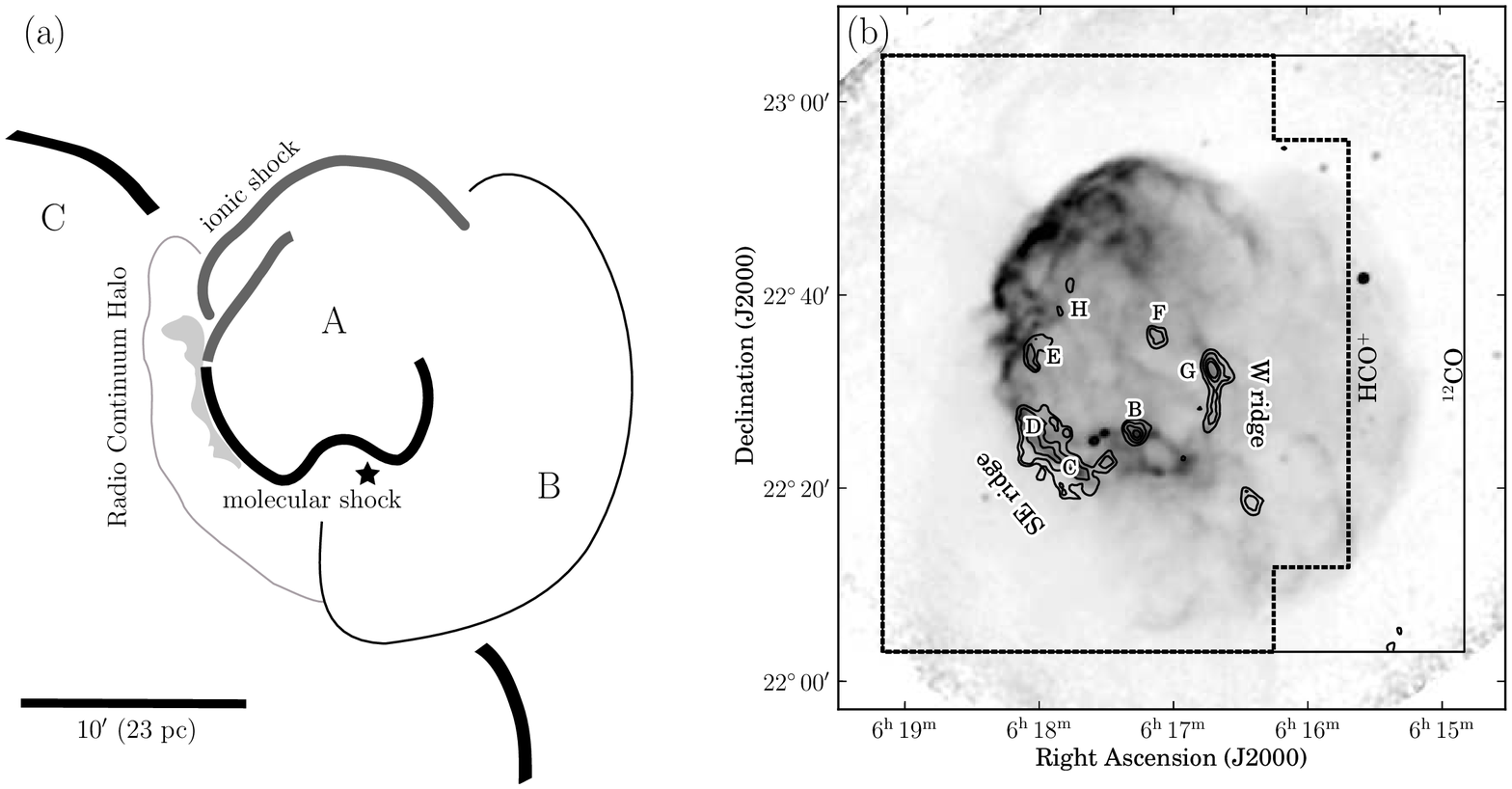}
\epsscale{1.}
\caption{(a) Schematic showing the overall morphology of
  \icfourfourthree\ adopted from \paperone, with the nomenclatures for
  shells overlaid. The asterisk indicates the position of the pulsar
  \citep{2001ApJ...554L.205O}.  (b) 21cm radio continuum image of
  \icfourfourthree\ (\paperone) with the area covered by the FCRAO
  observations: thin solid polygon for \twelveco\ ($J$=1$-$0) and thick
  dotted polygon for \hcop\ and \hcn. The contours are distribution of
  shocked \twelveco\ (see Fig.~\ref{fig:HCO+-pv} for
  details). The shock clumps (B--H) identified by
  \citet{1979ApJ...232L.165D} and \citet{1992ApJ...400..203D} are indicated.
  \label{fig:fcrao-field}}
\end{figure*}

Among the known 274 Galactic SNRs \citep{2009BASI...37...45G}, a few
tens of them show direct or indirect evidence of interaction with
ambient molecular clouds \citep[see][]{2010ApJ...712.1147J}.
\icfourfourthree\ (G189.1+3.0) is one of the first and most studied
SNRs for its interaction with molecular clouds.  The overall
morphology of \icfourfourthree\ is depicted in
Fig.~\ref{fig:fcrao-field}(a).  \icfourfourthree\ has two shells with
different radii \citep[Shells A and B, nomenclature is adopted
from][]{1986A&A...164..193B}, that are clearly seen in optical and
radio continuum images.
This peculiar morphology has led to a suggestion that the remnant
initially evolved in an inhomogeneous medium. \citet[][\paperone\
hereafter]{ic443paper1} suggested that Shell B could be a blowout part
of the remnant into a rarefied medium.  There is another larger and
fainter shell (Shell C) that partially overlaps with Shells A and
B. \citet{1986A&A...164..193B} proposed that Shell C is also physically
associated with Shells A and B, however, the work of
\citet{1994A&A...284..573A} suggests that Shell C is more likely another
SNR that is only positionally coincident in the sky
.

\icfourfourthree\ is apparently interacting with both atomic and
molecular gas (\paperone, and references therein). The SNR shock in the
northeastern part mostly shows the atomic lines expected in postshock
recombining gas \citep{1980ApJ...242.1023F}, and the shock is
propagating into an atomic medium. On the other hand, the southern
part of the remnant shows shock-excited broad molecular lines, and the
SNR shock is encountering molecular clouds
\citep{1979ApJ...228L..41D,1979ApJ...232L.165D,
1988MNRAS.231..617B,1989ApJ...341..857Z,1992ApJ...400..203D,1993A&A...279..541V}.
The Two Micron All Sky Survey \emph{JHK} composite image of \citet{2001ApJ...547..885R}
well demonstrates the different nature of the ambient medium.

Shocked molecular gas associated with \icfourfourthree\ was first
discovered in OH and \twelveco ($J$=1$-$0) by
\citet{1979ApJ...228L..41D,1979ApJ...232L.165D}.  Shock acceleration
is directly indicated by the large velocity width of the observed
lines ($\gtrsim 30\ \kms$).  Broad wings of various molecular lines
have been reported
\citep[e.g.,][]{1989ApJ...341..857Z,1993A&A...279..541V} including
H$_2$O \citep{2005ApJ...620..758S}.  The overall distribution of
shocked molecular gas around \icfourfourthree\ was mapped by
\citet{1988MNRAS.231..617B} in \htwo, by
\citet{1992ApJ...400..203D} in \twelveco\ ($J$=1$-$0) and \hcop\ ($J$=1$-$0),
and by \citet{2011ApJ...727...81X} in \twelveco\ ($J$=2$-$1, $J$=3$-$2).
Shocked
molecular emission is bright along the southern boundary of Shell~A,
but is also found toward the northeast and near the center of Shell~A
(Fig.~\ref{fig:fcrao-field}(b)).
No shocked molecular gas outside Shell
A has been reported. The overall distribution of shocked molecular gas
has often been described as an expanding ring
\citep{1988MNRAS.231..617B,1992ApJ...400..203D,1993A&A...279..541V}.

On the other hand, the identification and the nature of ambient clouds
that are physically interacting with \icfourfourthree\ have not been
comprehensively studied.
Ambient molecular clouds toward \icfourfourthree\ were first reported
by \citet{1977A&A....54..889C}. They found a geometrically thin ($\sim
3$ pc), sheet-like molecular clouds centered around $\vlsr \sim -3\ \kms$
and proposed its interaction with \icfourfourthree.
\citet{1992ApJ...400..203D} published more refined maps of the ambient
molecular gas.  \icfourfourthree\ is located toward the Galactic
anticenter direction where the radial velocity due to the Galactic
rotation is degenerate to $\vlsr \sim 0\, \kms$.  The clouds at
$\vlsr \sim -3\ \kms$ partially overlap with \icfourfourthree\ in the
sky and they are often assumed to be the clouds that are interacting
with \icfourfourthree\ \citep[e.g.,][]{2006ApJ...649..258T},
although no clear indication of their physical
association has been found.

Identifying the ambient clouds that are physically associated with
SNRs is important as it provides fundamental information about the
environment of the SNRs. This is particularly important for
core-collapse SNRs whose environment is significantly affected by
stellar feedback.
Also, the study of ambient clouds is important in understanding the
hadronic nature of associated $\gamma$-ray emission. $\gamma$-ray
emission can be emitted when cosmic rays that are accelerated at SNR
shocks encounter nearby dense molecular clouds.
\icfourfourthree\ has long been suspected being a site of cosmic ray
acceleration. It harbors an EGRET source 3EG~J0617+2238
\citep{1999ApJS..123...79H}, which might be associated with the hard
X-ray source in the eastern boundary \citep{2008ApJ...676.1050B}.  A
TeV-source in the western boundary is detected by MAGIC
\citep{2007ApJ...664L..87A} and VERITAS
\citep{2009ApJ...698L.133A}. More recently, GeV emission from
\icfourfourthree\ is reported by \emph{Fermi} LAT
\citep{2010ApJ...712..459A} and \emph{AGILE}
\citep{2010ApJ...710L.151T}.

In this paper, we report mapping observations of molecular lines toward the
Galactic SNR \icfourfourthree.
A square-degree field
toward \icfourfourthree\ is mapped in \twelveco($J$=1--0) and \hcop($J$=1--0) with
spatial resolution of $\sim 50\arcsec$.
Although \icfourfourthree\ has been observed in various molecular
lines, observations have been focused on the broad--line shocked
molecular clumps.
Our data present a global view of \icfourfourthree\ and its interaction
with molecular clouds.
The \twelveco\ observations we report here
were conducted using the Five College Radio Astronomy Observatory
(FCRAO) 14m radio telescope, the same telescope used by
\citet{1992ApJ...400..203D}.  \citeauthor{1992ApJ...400..203D} mapped
\icfourfourthree\ with full beam spacing (i.e., undersampled), but we
obtained fully sampled map using on-the-fly (OTF) technique.
The details of observations are described in \S~\ref{sec:obs}.  Our
observations find multiple components of ambient molecular clouds
toward \icfourfourthree. The characteristics of these
multiple cloud complexes are presented in
\S~\ref{sec:result}.
In \S~\ref{sec:nature-clouds-inter}, we try to identify ambient clouds
that are physically associated with \icfourfourthree.
The implication of our results on the environment of
\icfourfourthree\ is discussed in \S~\ref{sec:discuss} and
\S~\ref{sec:summary} summarizes the paper.

\section{Observations}
\label{sec:obs}

A $\sim 1^\circ\times 1^\circ$ area around \icfourfourthree\ was
observed in \twelveco\  $J$=1--0 (115.2712 GHz) and \hcop\
$J$=1--0 (89.18852 GHz).
All observations were obtained using the FCRAO 14 m telescope during 2003
January.  The single sideband 32 element focal plane array receiver
SEQUOIA was used in OTF mapping mode.  The dual channel
correlator (DCC) allowed us to observe two frequencies simultaneously.
An area of $60\arcmin \times 62\arcmin$ centered at (\hms{06}{17}{00},
\dms{22}{34}{00}) was mapped (Fig.~\ref{fig:fcrao-field}) with both
channels of DCC set to \twelveco\ but at different \vlsr, one at
$\vlsr = -45$ \kms\ and the other at $35$ \kms.  A smaller area of
$48\arcmin \times 61\arcmin$ centered at (\hms{06}{17}{26},
\dms{22}{34}{08}) was mapped with two channels of DCC set to \hcop\
and HCN at $\vlsr = -15$ \kms, respectively.
In this paper, however, we focus on the \twelveco\ and \hcop\ data as
the interpretation of the HCN data is complicated due the complexity
of its hyperfine structure.
The FWHM beam-size of the FCRAO telescope is $\sim 45\arcsec$
at 115 GHZ and $\sim 55\arcsec$ at 89 GHz.
The OTF data are regridded on a regular grid of 15$\arcsec$ pixels.
The DCC has a bandwidth of 50 MHz and 1024 spectral channels per IF
channel, leading to velocity resolution after Hanning smoothing of
$0.254$ and $0.328$ \kms\ at 115 GHz (CO) and 89 GHz (\hcop),
respectively.
For \twelveco, the velocity coverages were $-110\ \kms$ to $+20\ \kms$,
and $-30\ \kms$ to $+100\ \kms$ for each channel of the DCC.
Previous studies detected emission extending up to $+30\ \kms$\ in the
positive velocity and down to $-100\ \kms$ in the negative velocity
\citep[][and references therein]{2005ApJ...620..758S}.  Our
observations may not provide adequate velocity coverage for this very high
velocity gas.
We find that the lack of baseline is severe in some locations for
the positive velocity DCC channel data, while negligible for the
negative velocity DCC channel data.
As the negative DCC channel data cover virtually all the interesting
velocity range (see Section~\,\ref{sec:result}), we primarily use the
negative channel data.  For \hcop, the velocity coverage was $-100\
\kms$ to $+70\ \kms$.  The characteristic rms noise of each spectral
channel is $\sim0.2$ K for \twelveco, and $\sim0.5$ K for \hcop.
All the plotted data in this paper
are in antenna temperature ($T_{\mathrm{A}}^{*}$), which may be converted to
brightness temperature by dividing by the main beam efficiency which is
0.45 and 0.50 at 115 GHz and 90 GHz, respectively.
In Fig.~\ref{fig:fcrao-field}, we overlay
the areas we observed on top of the 21 cm radio continuum
image of \icfourfourthree\ (\paperone).

\section{Results}
\label{sec:result}

\begin{figure}[b]
\epsscale{0.8}
\plotone{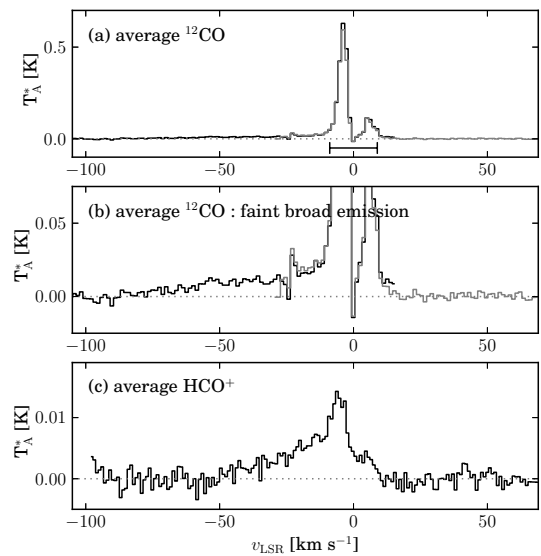}
\epsscale{1.}
\epsscale{1.}
\caption{(a) Average \twelveco\ ($J$=1--0) spectra of entire field
  observed. Two spectra, one centered at \vlsr=$-45$\ \kms\ and the
  other at \vlsr=35\ \kms, are shown. The velocity range for channel maps in
  Fig.~\ref{fig:fcrao-co-channel-map-a} is marked.
  The bump
  around $-25$ \kms\ is an artifact  due to an incomplete sky
  subtraction.
  (b) Same as (a) but $y$-range is adjusted
  to show the faint broad lines.
  (c) Average \hcop\ spectra of entire field observed.
  Note that the temperature scale is different.
\label{fig:fcra-co-mean-spec}}
\end{figure}

The spectra of \twelveco\
and \hcop\ lines averaged over the entire observed field are
shown in Fig.~\ref{fig:fcra-co-mean-spec}.
Most of the CO emission arises from two narrow velocity components at
approximately $-3$ \kms\ and $+5$ \kms, presumably from ambient
molecular cloud. However very weak and broad
emission from the shocked gas can also be seen. The average \hcop\
spectra are dominated by broad lines from the shocked gas.
The average spectrum of \twelveco\ in
Fig.~\ref{fig:fcra-co-mean-spec} shows an artifact
around \vlsr = $-20$ \kms, which is the velocity of
atmospheric line and we attribute this artifact to an imperfect sky
subtraction. This artifact is only seen in
the average spectrum and has negligible
effect on individual \twelveco\ spectra.

\subsection{Shocked Molecular Gas}
\label{sec:shock-molec-gas}

\begin{figure*}
\epsscale{1.}
\plotone{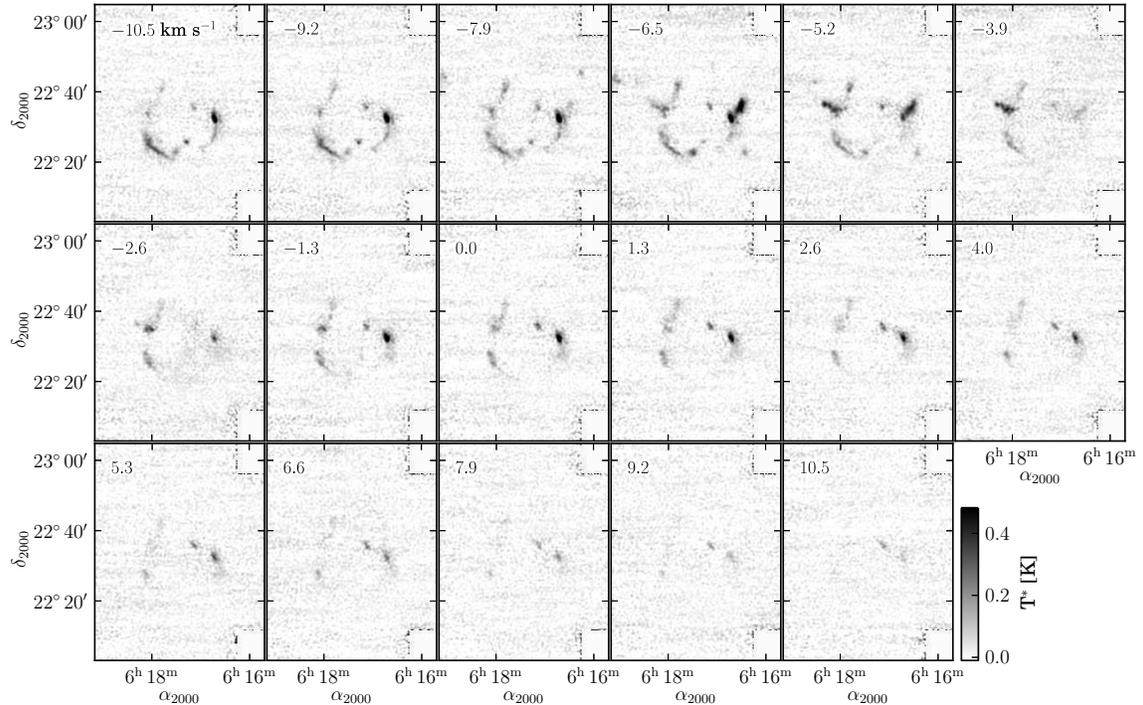}
\epsscale{1.}
\caption{\hcop\ channel maps for velocities between $-10$ \kms\ and
  $+10\ \kms$. The emission is predominantly from shocked molecular
  gas.
  However, there is some contribution from the ambient clouds, as can
  be identified by narrow lines in the PV map (Fig.~\ref{fig:HCO+-pv}).
  \label{fig:HCO+-channelmaps}}
\end{figure*}

\begin{figure*}
\epsscale{.4}
\centering{
\includegraphics[scale=0.8]{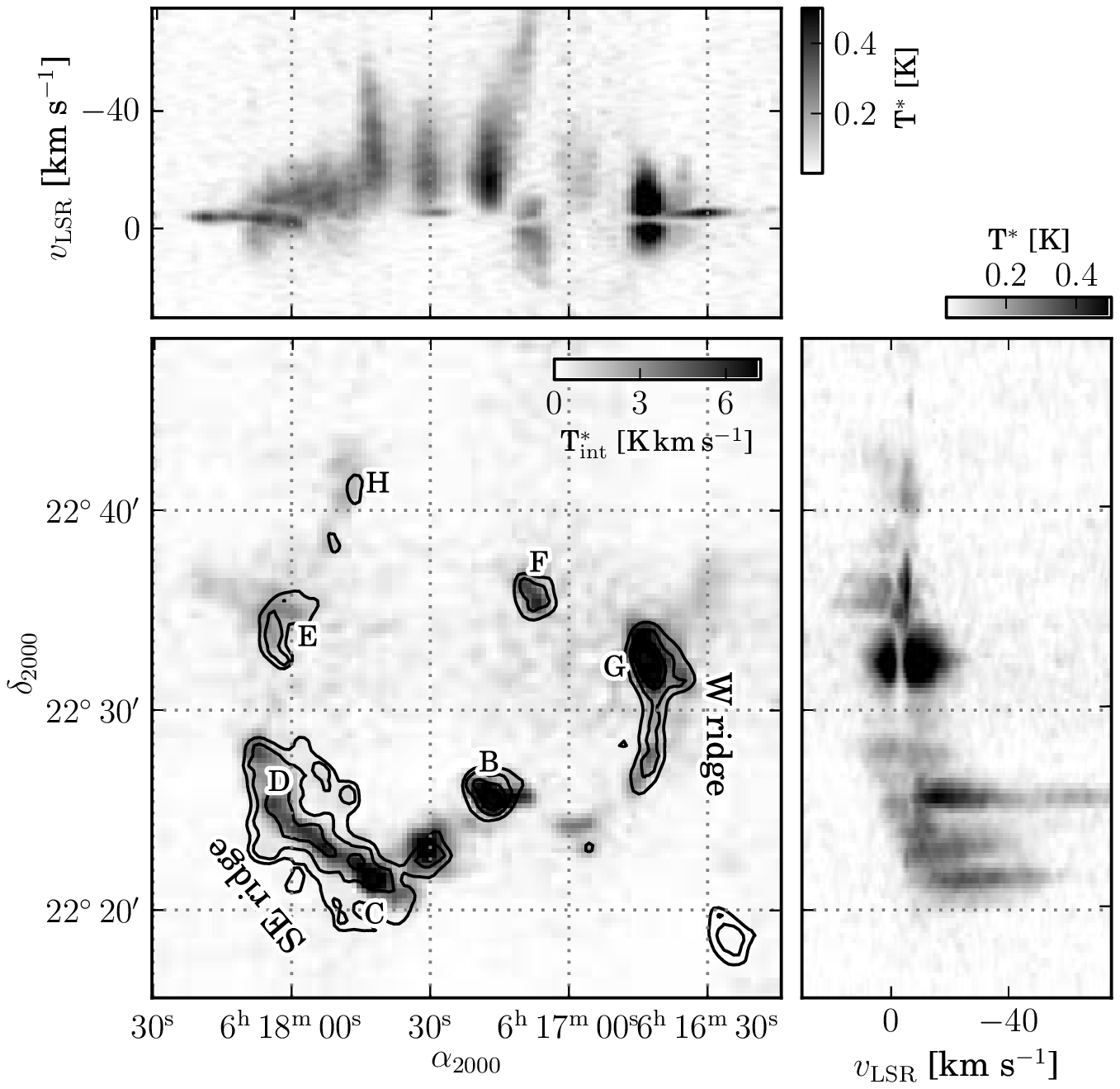}
}
\epsscale{1.}
\caption{ \hcop\ integrated temperature map on the lower-left panel, and
  position-velocity maps on the upper and right panels. The upper
  panel is an R.A.--velocity map of peak temperatures along the decl.\
  axis. The right panel is a velocity--decl.\ map of peak temperatures
  along the R.A.\ axis. The values are derived from the whole
  velocity range.
  The \hcop\ integrated temperature map has an rms error of 0.3
  K$\,\kms$. The R.A.--velocity map and the velocity--decl.\ map has an
  rms error of $0.03$ K.
  The contours represent \twelveco\ integrated
  temperature map over velocity ranges of $-17$ to $-10$ \kms\ and $+1$ to
  $+3$ \kms, chosen to minimize the contribution
  from the ambient gas.  The contour levels are 1.2, 2.6, 6, 10
  K$\,\kms$ and the rms error is $0.3$ K$\,\kms$.
  The shocked
  clumps are also indicated as in Fig.~\ref{fig:fcrao-field}(b).  Some
  \hcop\ emission is from the ambient gas as can be identified from
  their narrow line widths. These ambient gas components are
  associated with the small clouds (see
  \S~\ref{sec:small-isolated-clouds}).
\label{fig:HCO+-pv}}
\end{figure*}

The shocked molecular gas is characterized by its broad line wings
\citep[see][and references therein]{2005ApJ...620..758S}.
For \vlsr\ $< -10$ \kms\ and \vlsr\ $> +10$ \kms, the emission is
confined spatially to small regions with linewidths that are very
broad. This emission is assumed to arise almost entirely from the
shocked gas. Due to projection effects, the velocity of some
of the shocked gas is close to 0 \kms\ \citep{1992ApJ...400..203D} and
a significant fraction of the shocked gas is indeed in the
velocity range between \vlsr $= -10$ to $+10$ \kms.
But the
study of this low velocity shocked gas with \twelveco\ is
difficult due to the contamination by the ambient gas
emission
(Fig.~\ref{fig:fcra-co-mean-spec}).
\hcop\ emission, on the
other hand, has less contamination from the ambient gas.
The channel maps and the position-velocity (PV) maps of the observed \hcop\ line are shown in
Fig.~\ref{fig:HCO+-channelmaps} and in Fig.~\ref{fig:HCO+-pv}.
In particular, the PV maps
clearly show that most of the emission has a line width $\gtrsim
10\ \kms$, and thus is from the shocked gas. Some ambient gas is
picked up as absorption features and also as emission features.
Thus \hcop\ is better suited for studying the overall distribution of
the shocked molecular gas.
In Fig.~\ref{fig:HCO+-pv}, the distribution of high-velocity
\twelveco\ emission is compared to the distribution of \hcop\
emission.
We note that the \twelveco\ image at higher velocity can provide more
sensitive and detailed distribution of shocked gas compared to \hcop\
data with a caveat that the lower velocity shocked gas is not
included.

The distribution of shocked molecular gas takes a shape of an
incomplete ring, which consists of several bright clumps and fainter
emission that seems to connect these clumps.
The shocked gas emission is brighter in the
south,
while the emission in the north is weaker.
There is a
general tendency of the systematic velocity of shocked gas
getting more positive with
the increasing declination.
This large-scale velocity gradient is consistent with previous studies
which suggested that the
large-scale distribution of the shocked gas in \icfourfourthree\
could be described as an
expanding ring \citep{1988MNRAS.231..617B,1992ApJ...400..203D}.
On the other
hand, we note that long ($\gtrsim$ a few arcminutes) filaments of
shocked gas are only prominent in the southeastern (SE ridge) and
western parts (W ridge) of the proposed ring
(Figure~\ref{fig:HCO+-pv}) and the gas kinematics are
dominated  by rapid
variation on the scale of $\lesssim$ a few arcminutes, which are
referred as shocked ``clumps''.
The PV plot in right ascension (the top panel of Fig.~\ref{fig:HCO+-pv})
shows a tilt of the emission in clumps B and C.
In clump B, the emission shifts to smaller R.A. at higher shock
velocities, while for clump C, the emission shifts to larger R.A. at
higher shock velocities.  Little shift is seen in the PV plot in
declination.  The velocity shift seen in Clump B was first detected by
\citet{1992ApJ...400..203D} in both \twelveco\ and \hcop\ emission.
They suggested that this shift was a sign that the gas was being
ablated and accelerated away from the shocked clump.  It is
interesting that we see a similar effect in Clump C, although the
ablated gas is moving off in the opposite direction.  Although the
geometry of the interaction of the SNR with the ambient gas is
complicated, these tilts in the PV diagram provide an indication of
the direction of the shock.

\label{sec:result-hcop}

\label{sec:hcop-result-shocked}

\subsection{Ambient Molecular Gas}
\label{sec:ambi-molec-clouds}

\begin{figure*}
\epsscale{0.7}
\plotone{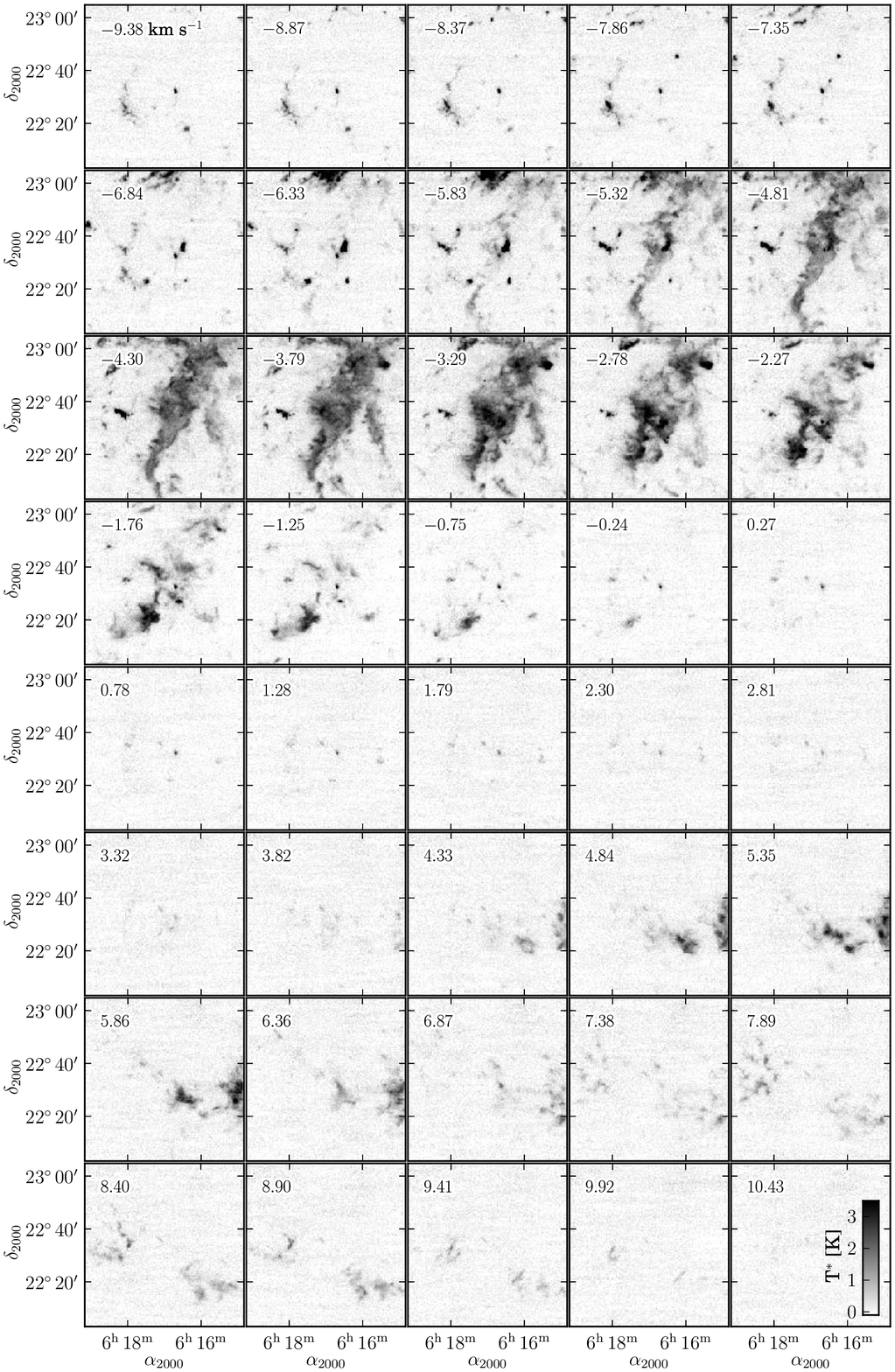}
\epsscale{1.}
\caption{\twelveco\ channel maps for velocities between $-10$ \kms\
  and $+10\ \kms$, mostly tracing the ambient molecular gas.
  Each panel represents mean temperature for $\Delta v = 0.5\
  \kms$, where their central velocity is shown in the top-left corner.
  \label{fig:fcrao-co-channel-map-a}}
\end{figure*}

The channel maps of \twelveco\ in the velocity range of \vlsr =
$-10$~\kms\ to $+10$~\kms, where most of the emission traces
ambient clouds, are shown in Fig.~\ref{fig:fcrao-co-channel-map-a}.
 \icfourfourthree\ is
located toward the Galactic anticenter direction,
and no ambient molecular gas emission is found outside of this
velocity range.  The channel maps in
Fig.~\ref{fig:fcrao-co-channel-map-a}, together with the mean
\twelveco\ spectrum in Fig.~\ref{fig:fcra-co-mean-spec}, suggest that
there are two distinct complexes of molecular gas, one in the negative
velocity of $-10\ \kms < \vlsr < 0\ \kms$ and the other in the
positive velocity of $+3\ \kms < \vlsr < +10\ \kms$.

\label{sec:co-result-ambient}
\label{sec:result-diffuse-cloud}

\begin{figure}
\epsscale{0.45}
\plotone{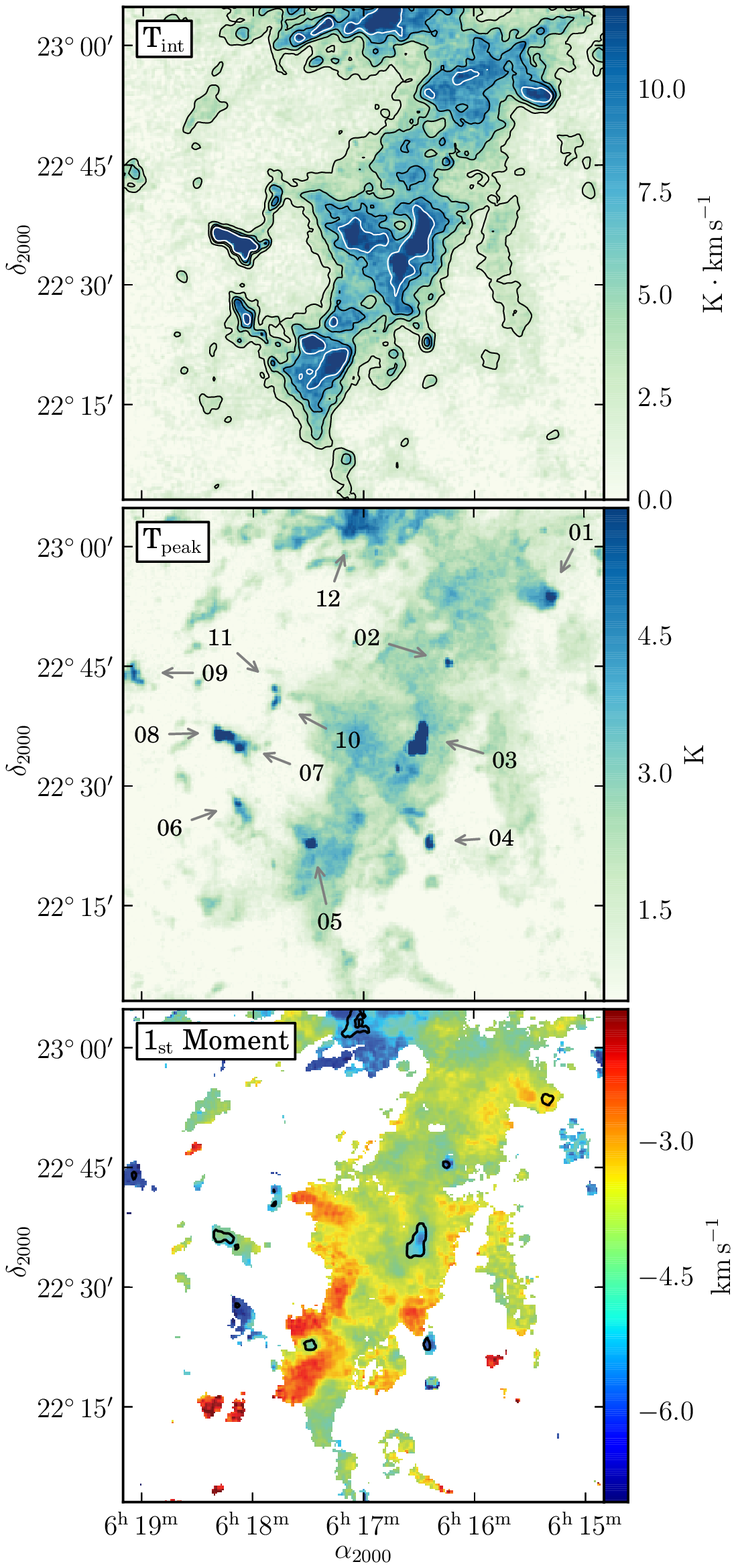}
\epsscale{1.}
\caption{
(a) \twelveco\ integrated temperature map in velocity range of
$-10$ to $0\ \kms$, mostly tracing the ambient gas.
(b) \twelveco\ peak temperature map over the same velocity
interval.
The numbers denote the location of small
clouds (\SICs) discussed in \S~3.3.
(c) The intensity
weighted mean velocity
(first moment) of the \twelveco\ emission.
The contours mark the regions which satisfy our selection
criteria for \SICs\ (see \S~\ref{sec:small-isolated-clouds} for details).
\label{fig:int_peak_firstmom}}
\end{figure}

\begin{figure*}
\epsscale{0.6}
\centering{
\includegraphics[scale=0.8]{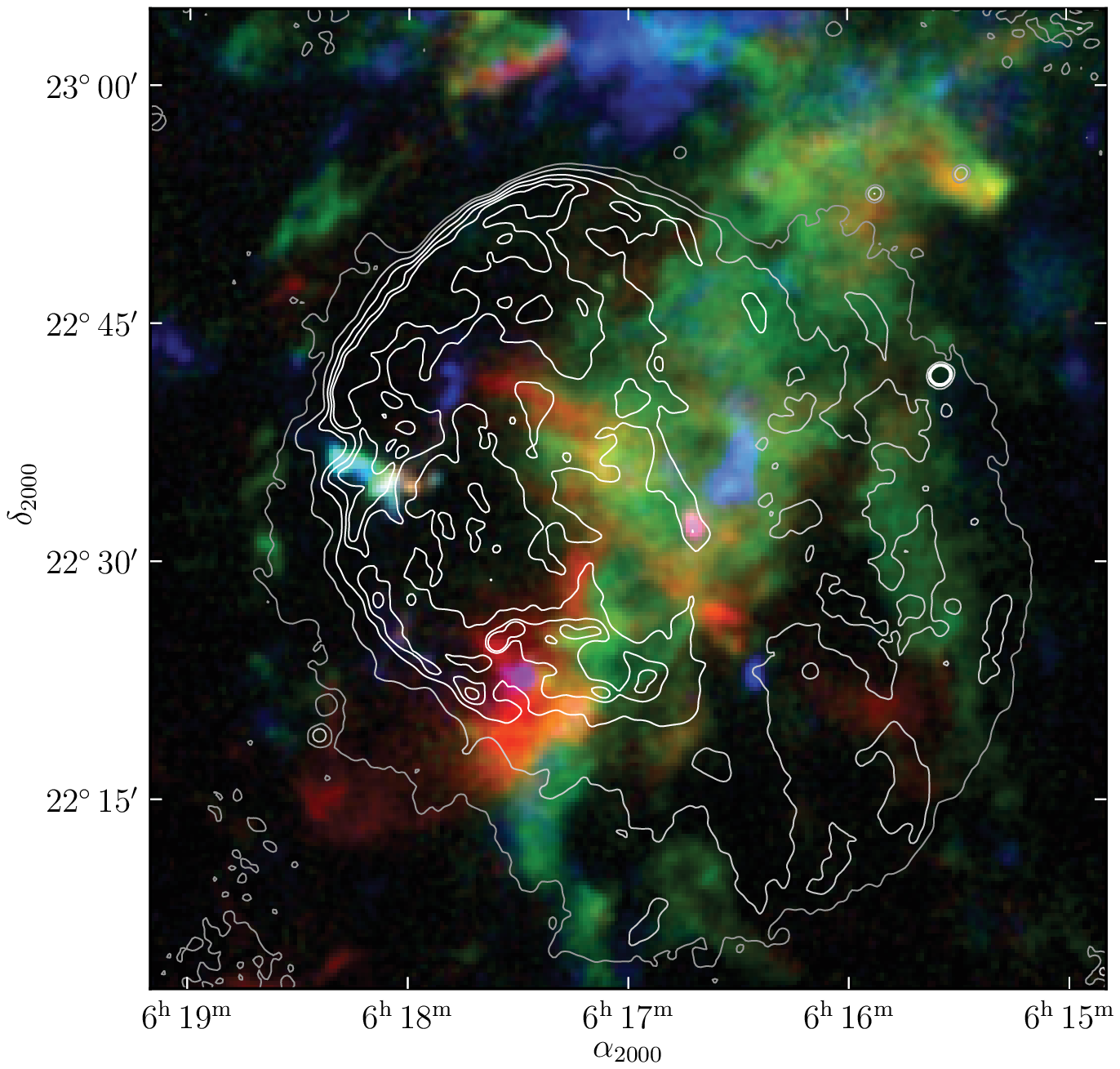}
}
\epsscale{1.}
\caption{RGB composite image
where each  R, G, and B channel  corresponds
to \twelveco\ integrated temperature maps of velocity ranges $-0.5$ to
$-3.4$, $-3.4$ to $-5.1$ and $-5.1$ to $-7.0\ \kms$, respectively.
The contours are 21cm radio continuum image from \paperone. The
contour levels are 10, 20, 40, 60, 80 mJy.
\label{fig:CO-RGB}}
\end{figure*}

\begin{figure*}
\epsscale{0.6}
\centering{
\includegraphics[scale=0.8]{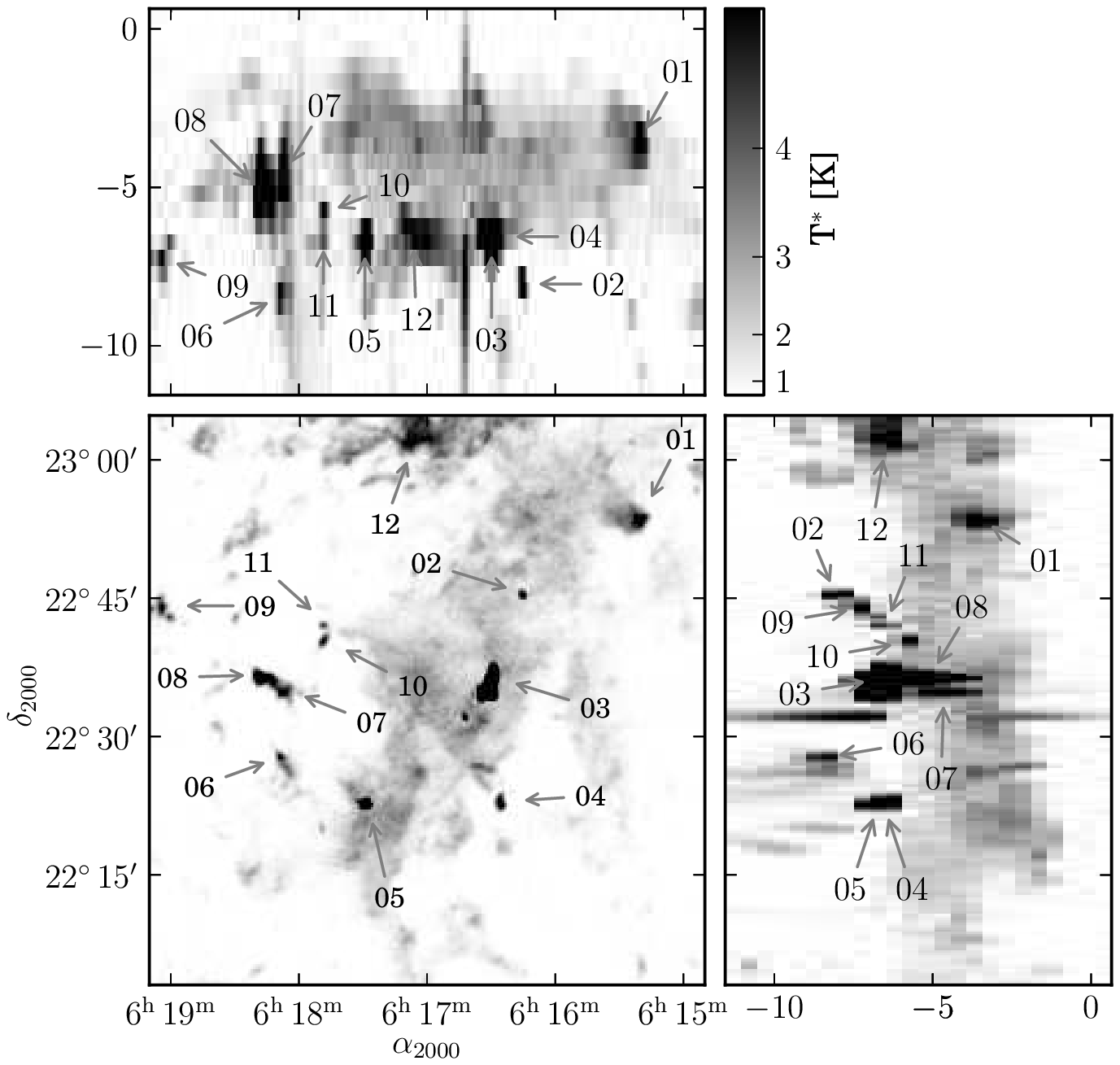}
}
\epsscale{1.}
\caption{\twelveco\ peak temperatures map (identical to
  Fig.~\ref{fig:int_peak_firstmom}(b)) together with position--velocity
  maps. The upper panel is a R.A.--velocity map of peak temperatures
  along the decl.\ axis. The right panel is
  a velocity--decl.\ map of peak temperatures along the R.A.\ axis.
  The images are shown in a squared scale to better visualize the
  locations of the \SICs.
  The numbers are identification numbers for \SICs. The central
  velocities of
  \SICs\ are adopted from $v_0$ of Component~1 in Table~\ref{tbl-peak-t-clumps}.
  \label{fig:CO-PV-Tpeak}}
\end{figure*}

The clouds at negative velocities
show significant channel-to-channel variation.
To get a grasp of the overall structure of the cloud complex, we
generate an integrated temperature map, a peak temperature map and a
first moment (intensity weighted velocity) map of \twelveco\ emission
in the velocity range $-10 < \vlsr < 0\ \kms$, which are shown in
Fig.~\ref{fig:int_peak_firstmom}.  Also, in Fig.~\ref{fig:CO-RGB}, an
RGB composite image is shown, where each red, green and blue channel
corresponds to an integrated temperature map of different velocity
ranges (see the figure caption for more details).
The integrated temperature map provides an overall view of the spatial
distribution of
this gas, although there is some contamination by emission from the
shocked gas.
Most of the ambient gas is distributed along the
northwest-southeast direction across the field, roughly corresponding to
the interface between Shells A and B (Fig.~\ref{fig:CO-RGB}).
The peak temperature map is similar to the integrated temperature map,
however, several small clouds ($\sim 1\arcmin$) with relatively higher
peak temperatures ($\gtrsim$\ 5 K) stand out against the diffuse
clouds whose peak temperatures are generally less than 3 K.  These
small clouds are even more prominent in the PV
maps in Fig.~\ref{fig:CO-PV-Tpeak}. Fig.~\ref{fig:CO-PV-Tpeak} clearly
reveals two distinguished kinematic characteristics of the small
clouds compared to the diffuse clouds. First, the linewidths of
these small clouds are much narrower than the diffuse
clouds. Second, most of these small clouds show clear offset in velocity
relative to the diffuse emission.  Although some of these small clouds
are located toward the diffuse clouds, a significant velocity
offset between them suggests that most of the small clouds are not
related to the diffuse clouds.
Thus, these small clouds are considered to be
separate clouds from the diffuse clouds.
For the purpose of the discussion, we denote the
diffuse clouds between $-6\ \kms$ and $-1\ \kms$ as
``\minusthreecloud'' and the small high brightness temperature clouds
as ``\SICs''. The characteristics of the \SICs\ will be further
discussed in \S~\ref{result-small-clouds}.
The first moment map mostly traces the radial velocity variation
within the \minusthreecloud\ and suggests that the main body of the
cloud has a small, systematic velocity shift from east to west. A
similar tendency can be noticed in the RGB composite image
(Fig.~\ref{fig:CO-RGB}).

\begin{figure}[t]
\epsscale{0.8}
\plotone{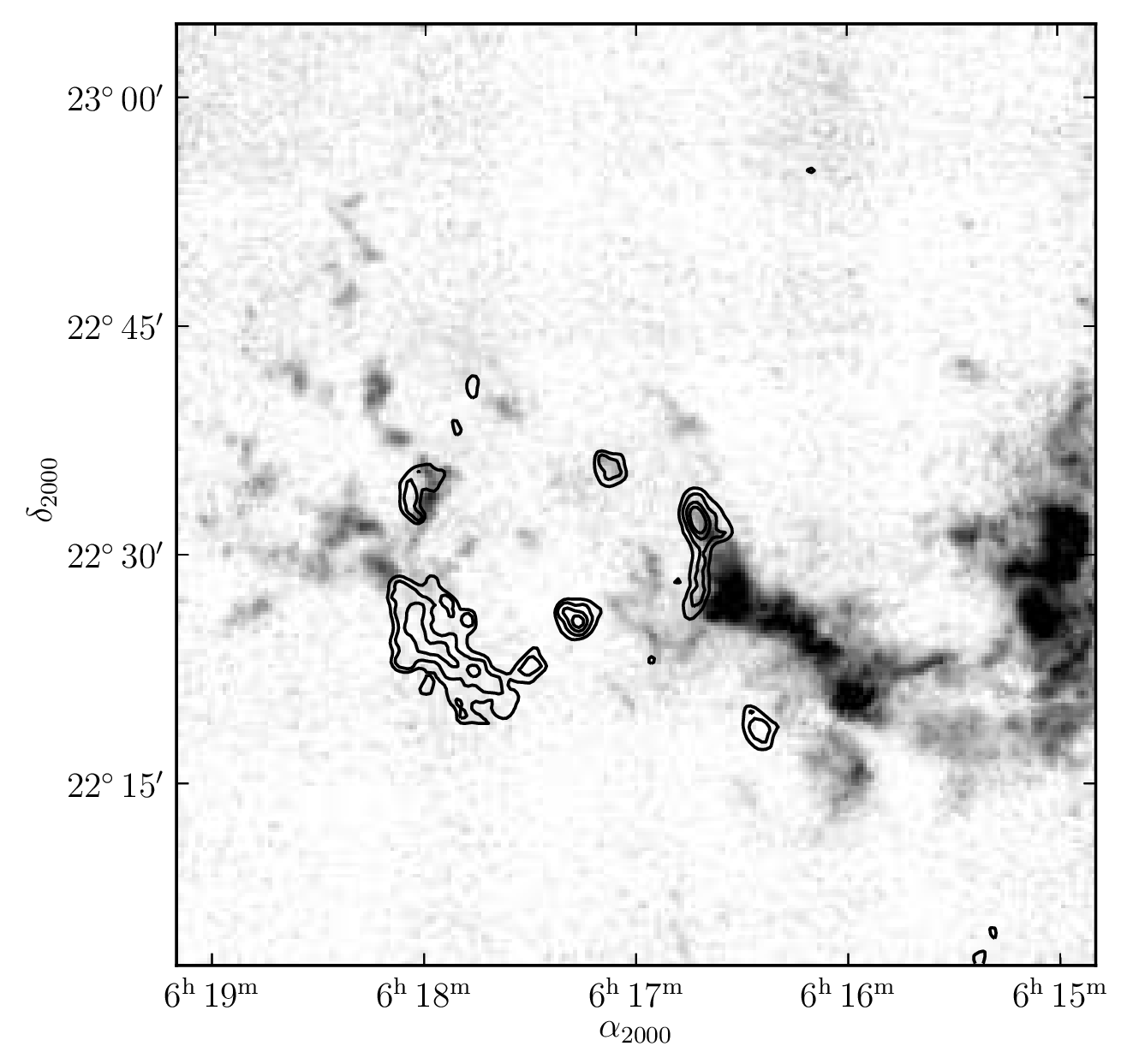}
\epsscale{1.}
\caption{\twelveco\ distribution of \plussixcloud, integrated over velocity
  range of $+4\ \kms$ to $+10\ \kms$.  The
  contours are \twelveco\ integrated temperature map as in
  Figure~\ref{fig:HCO+-pv}, tracing the distribution of shocked
  molecules.
  \label{fig:co-positive-inter}}
\end{figure}

In the positive velocity range,
relatively fainter emission from ambient clouds is seen between
$\vlsr=+3$ and $+10\ \kms$. We call these clouds as ``\plussixcloud''.
Emission
between $0\ \kms$ and $+3\ \kms$ is predominantly from shocked
molecular gas.
Fig.~\ref{fig:co-positive-inter} shows integrated
temperature map of $\vlsr=+4$ to $+10\ \kms$, which represent mostly
the ambient gas with some contribution of the shocked gas.
The clouds seem to be divided into two parts: the western part and the eastern part.
The western part is brighter and shows a prominent
filamentary structure. The eastern part is fainter and shows a number
of small clouds.
It is noteworthy that the western and the eastern parts are separated
by Shell A (Fig.~\ref{fig:co-positive-inter}).

\subsection{Small Clouds (\SICs)}

\label{result-small-clouds}
\label{sec:small-isolated-clouds}

\newcommand{\tpeak}{$T_{\mathrm{peak}}^{*}$}
\newcommand{\tint}{$T_{\mathrm{int}}^{*}$}
\begin{figure}
\epsscale{0.6}
\plotone{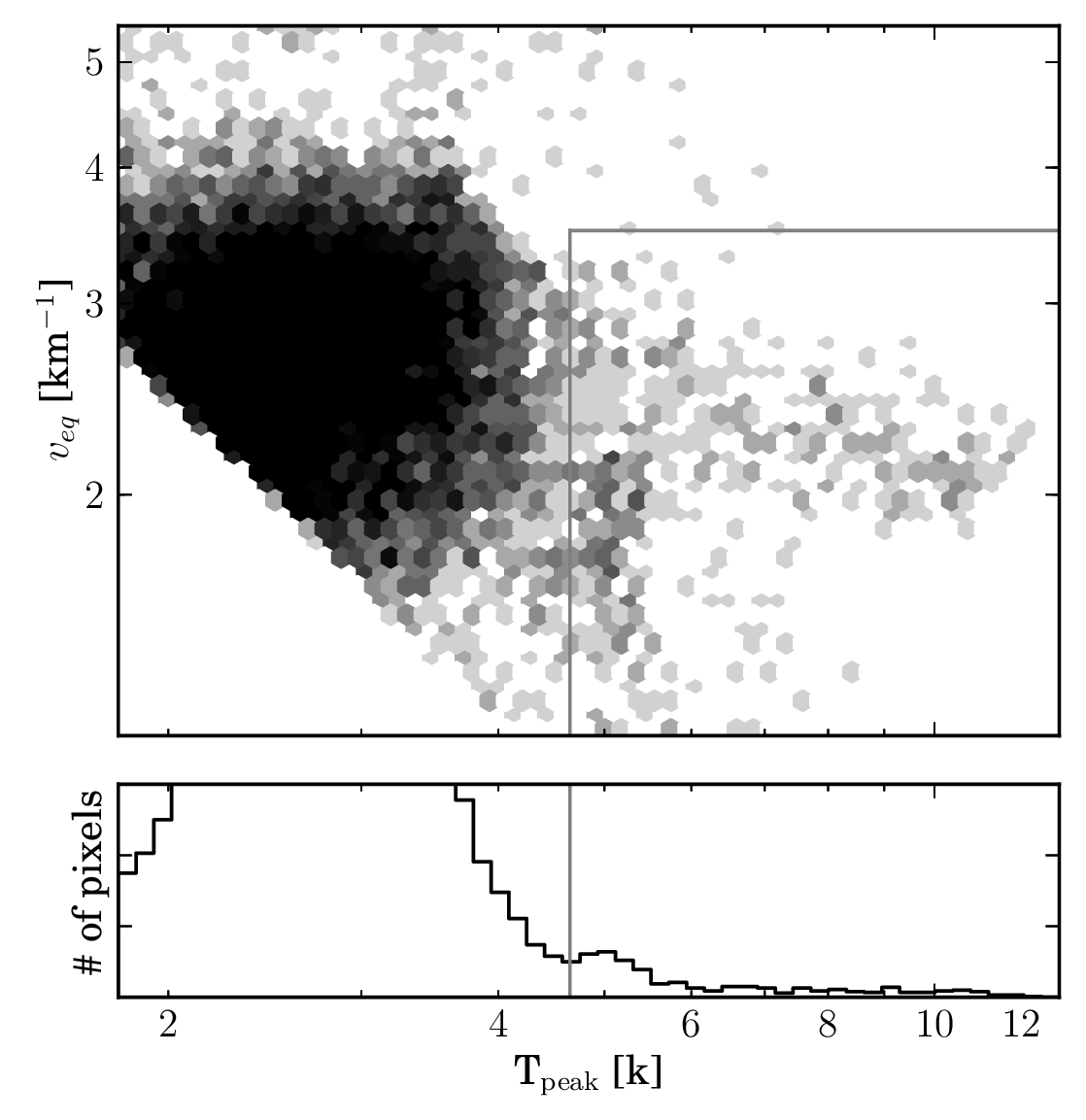}
\epsscale{1.}
\caption{(top) distribution of \tpeak\ vs.\ $\Delta v_{\mathrm{eq}}$
  (equivalent line width), where $\Delta v_{\mathrm{eq}}$, defined as
  \tint / \tpeak.  While most pixels have \tpeak $\lesssim 4$~K and
  $\Delta v_{\mathrm{eq}}$ $\sim$ 2 \kms, there are pixels with \tpeak
  $\gtrsim 5$~K and $\Delta v_{\mathrm{eq}} \lesssim$ 3 \kms\ which
  are associated with \SICs. The gray lines mark the selection
  criteria used to identify \SICs. (bottom) Histogram of \tpeak. A
  clear bump around 5~K is seen. The gray vertical line marks
  \tpeak=4.6 K above which are necessary condition to be identified as
  \SICs.
  \label{fig:CO-TT}}
\end{figure}

\begin{figure*}

\epsscale{0.8}
\centering{
\includegraphics[scale=0.8]{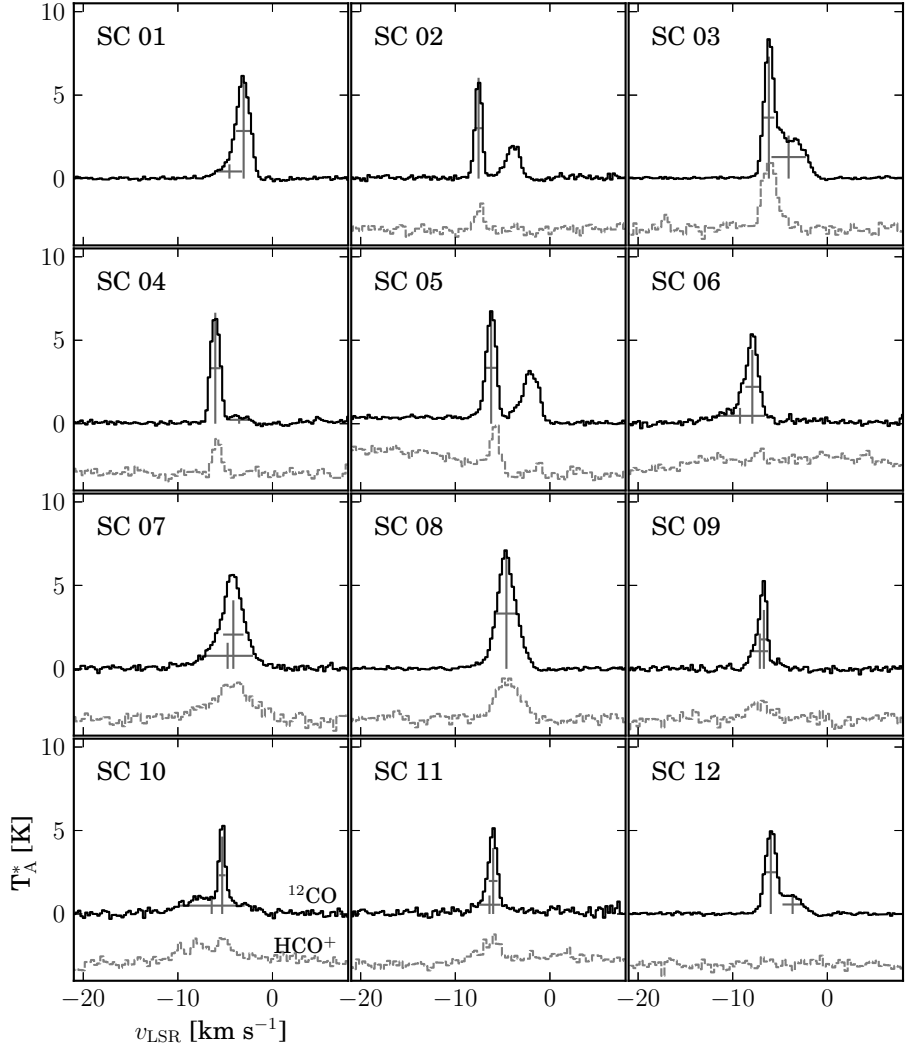}
}
\epsscale{1.}
\caption{\twelveco\ line profiles of the \SICs. The
  crosses indicate the line center and the full-width-at-half-maximum
  when fit by multiple Gaussian components (see
  Table~\ref{tbl-peak-t-clumps}). \hcop\ spectra at the same locations are
  also shown as gray lines. No \hcop\ spectrum is available for
  \SIC~01 as the observed area for \hcop\ is smaller than that for
  \twelveco.  \label{fig:clumps-line-profiles}}

\end{figure*}

\begin{figure*}

\epsscale{0.8}
\centering{
\includegraphics[scale=0.8]{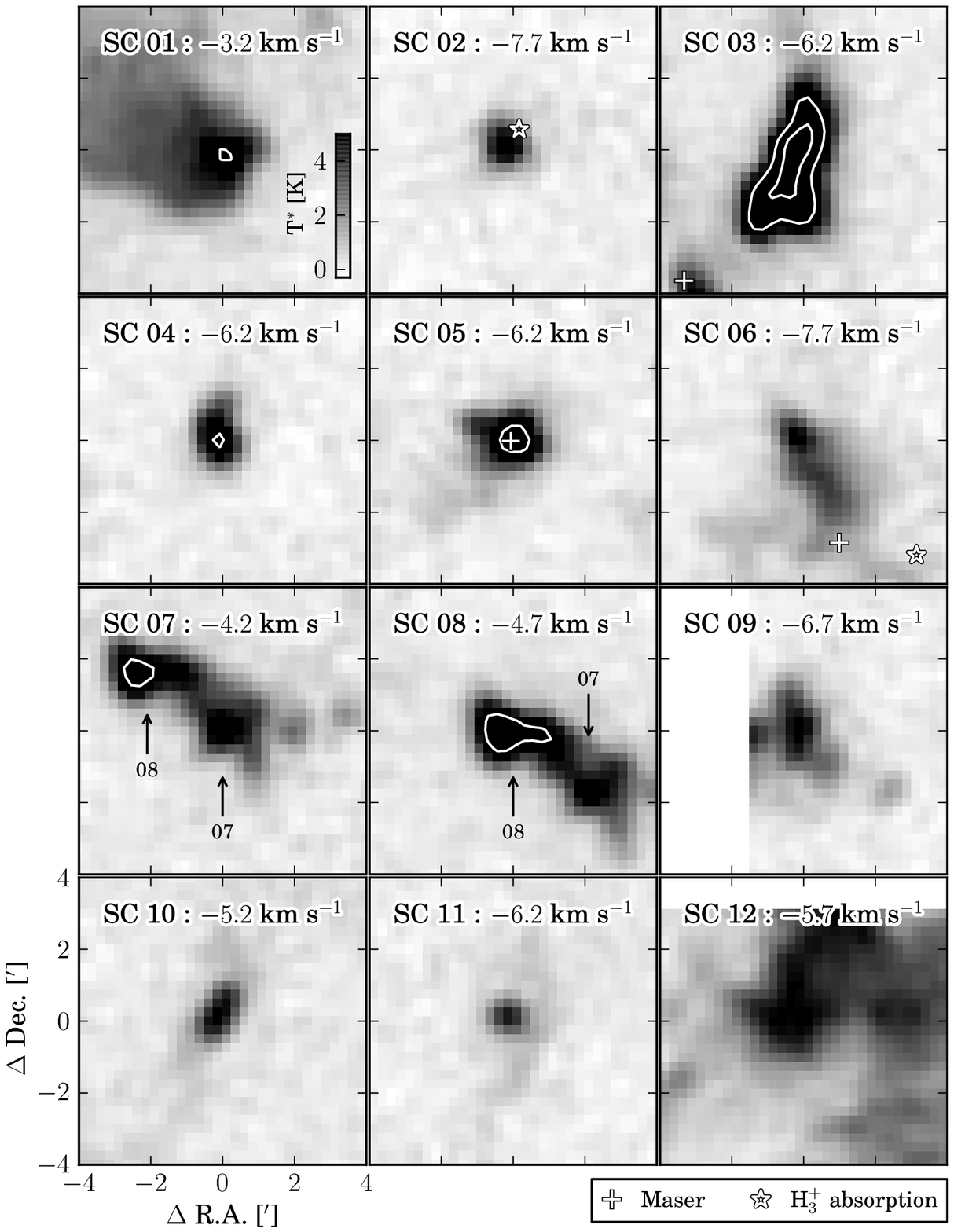}
}
\epsscale{1.}

\caption{$8\arcmin \times 8\arcmin$ close-up view of each \SIC\ at
  its peak velocity (see Table~\ref{tbl-peak-t-clumps}). Each
  panel is labeled with the \SIC\ number and the channel velocity
  used. The white contours are $\mathrm{T}^*=$ 5 K, 7.5 K and 10
  K. The plus signs near \SICs\ 03, 05 and 06 are positions of OH
  maser sources and the star signs near \SIC\ 02 and 06 are sight
  lines along which H$_3^+$ absorption features are detected (see
  \S~\ref{sec:isolated-clumps}).
  The locations of \SIC~07 and \SIC~08
  are also marked with arrows for clarification.
  \label{fig:clumps-zoom-up}}
\end{figure*}

The \twelveco\ peak temperature map in
Fig.~\ref{fig:int_peak_firstmom} and the PV maps in
Fig.~\ref{fig:CO-PV-Tpeak} reveal the existence of several SCs
($\sim1\ \arcmin$)  with relatively high brightness temperature
and narrow line width.
To identify these clouds in more systematic way, we utilize two
characteristic quantities: the peak temperatures (\tpeak) and the
equivalent line width ($\Delta v_{\mathrm{eq}}$).  $\Delta
v_{\mathrm{eq}}$ is defined as the ratio of the integrated
temperatures \tint\ to \tpeak. These temperatures are measured in the
velocity range of $-10$ \kms\ to $0$ \kms.
Fig.~\ref{fig:CO-TT} shows a two dimensional histogram of these two
quantities measured from the pixels in the observed area.  Most pixels
have \tpeak$\lesssim 4$~K and $1 \lesssim v_{\mathrm{eq}} \lesssim\ 4\
\kms$,
and we find that they are mostly associated with \minusthreecloud.
On the other hand, there are pixels with relatively high peak
temperature (\tpeak $\gtrsim$ 4.5~K) with relatively narrower
equivalent line width ($v_{\mathrm{eq}} \lesssim 3\ \kms$). We find these pixels
correspond to the \SICs\ as suspected. Based on the two-dimensional histogram, we
adopt a criteria of \tpeak $>$ 4.6~K and $v_{\mathrm{eq}} < 3.5\
\kms$, which results in the identification of 12 SCs,
and these are identified in
Fig.~\ref{fig:int_peak_firstmom}(b) and labeled with numbers (e.g.,
SC~01).
The selection process includes all of the SCs that would have
been chosen by eye, as we are primarily selecting them based on peak
temperature.

The \twelveco\ line profiles of the 12 \SICs\ are shown in
Fig.~\ref{fig:clumps-line-profiles}.
These are average spectra extracted from
the contours shown in Fig.~Fig.~\ref{fig:int_peak_firstmom}(b),
and they all show narrow lines with high peak temperatures as
expected.  We fit the spectra with Gaussian profiles and the
results are summarized in Table~\ref{tbl-peak-t-clumps}. Often, the
spectra need multiple Gaussian components to be adequately fit.
The additional components are contributed by emission from either the
diffuse clouds or the shocked gas.

Central velocities of these \SICs\ range between $-3$ and $-8\,\kms$,
with a median of $\sim -6\,\kms$.
Other than \SIC\ 01, 07/08 (see the note in Table~1), they have peak
velocities more negative than $-5\ \kms$, and are well separated from
\minusthreecloud\ in the velocity space (Fig.~\ref{fig:CO-PV-Tpeak}).
Their line widths are between
$1$ and $2\ \kms$.  The \hcop\ profiles are also plotted in
Fig.~\ref{fig:clumps-line-profiles} when available.  The \SICs\ tend
to have similar narrow emission lines in \hcop.
The \twelveco\ maps of $\sim 8\arcmin\times8\arcmin$ regions around
individual \SICs\ at their peak intensity channels are shown in
Fig.~\ref{fig:clumps-zoom-up}.
Part of the emission seen around \SICs\ 05, 06, 07/08, 10, and 11 is
due to shocked gas, and the association of these \SICs\ to the shocked molecular
gas will be further discussed in \S~\ref{sec:isolated-clumps}.

\ifaastex
\begin{deluxetable}{cccrrrcrrr}
\rotate
\else
\begin{deluxetable*}{cccrrrcrrr}
\fi
\tablecaption{Spectral parameters of the small clouds (\SICs; see \S~\ref{sec:small-isolated-clouds})\label{tbl-peak-t-clumps}}
\tablewidth{0pt}
\tablecolumns{10}
\tablehead{\colhead{}& \colhead{}& \colhead{}
& \multicolumn{3}{c}{Component 1}
& \colhead{}
& \multicolumn{3}{c}{Component 2} \\
\cline{4-6} \cline{8-10}
\colhead {\SIC\ No.} & \colhead {$\alpha$ (2000)} & \colhead {$\delta$ (2000)}
& \colhead {T$_{\mathrm{A}}^{*}$ [K]} & \colhead {$v_0$} & \colhead {$v_{\mathrm{FWHM}}$}
& \colhead{}
& \colhead {T$_{\mathrm{A}}^{*}$ [K]} & \colhead {$v_0$} & \colhead {$v_{\mathrm{FWHM}}$}
}
\startdata

01 &
\hmss{06}{15}{20}{1} & \dms{22}{53}{43}
&  5.7 $\pm$ 0.5 
& -3.1 $\pm$ 0.1  
& 1.6 $\pm$ 0.1  
& &  0.9 $\pm$ 0.3 
& -4.6 $\pm$ 0.5  
& 2.7 $\pm$ 0.6  
\\

02 &
\hmss{06}{16}{14}{4} & \dms{22}{45}{14}
&  6.0 $\pm$ 0.1 
& -7.5 $\pm$ 0.1  
& 0.9 $\pm$ 0.1  
& & 
& 
& 
\\

03 &
\hmss{06}{16}{29}{6} & \dms{22}{36}{14}
&  7.3 $\pm$ 0.1 
& -6.2 $\pm$ 0.1  
& 1.2 $\pm$ 0.1  
& &  2.6 $\pm$ 0.1 
& -4.1 $\pm$ 0.1  
& 3.6 $\pm$ 0.1  
\\

04 &
\hmss{06}{16}{25}{3} & \dms{22}{22}{44}
&  6.6 $\pm$ 0.1 
& -6.0 $\pm$ 0.1  
& 1.2 $\pm$ 0.1  
& &  0.5 $\pm$ 0.1 
& -3.5 $\pm$ 0.1  
& 2.0 $\pm$ 0.2  
\\

05 &
\hmss{06}{17}{29}{1} & \dms{22}{22}{44}
&  6.7 $\pm$ 0.1 
& -6.2 $\pm$ 0.1  
& 1.3 $\pm$ 0.1  
& & 
& 
& 
\\

06 &
\hmss{06}{18}{08}{1} & \dms{22}{27}{44}
&  4.4 $\pm$ 0.2 
& -7.9 $\pm$ 0.1  
& 1.5 $\pm$ 0.1  
& &  0.9 $\pm$ 0.1 
& -9.2 $\pm$ 0.2  
& 5.2 $\pm$ 0.3  
\\

07 &
\hmss{06}{18}{08}{2} & \dms{22}{34}{58}
&  4.1 $\pm$ 0.3 
& -4.1 $\pm$ 0.1  
& 2.2 $\pm$ 0.1  
& &  1.6 $\pm$ 0.1 
& -4.7 $\pm$ 0.1  
& 5.4 $\pm$ 0.3  
\\

08 &
\hmss{06}{18}{17}{9} & \dms{22}{36}{28}
&  6.6 $\pm$ 0.1 
& -4.6 $\pm$ 0.1  
& 2.4 $\pm$ 0.1  
& & 
& 
& 
\\

09 &
\hmss{06}{19}{04}{6} & \dms{22}{43}{57}
&  3.5 $\pm$ 0.4 
& -6.7 $\pm$ 0.1  
& 0.7 $\pm$ 0.1  
& &  2.1 $\pm$ 0.2 
& -7.1 $\pm$ 0.1  
& 1.7 $\pm$ 0.1  
\\

10 &
\hmss{06}{17}{48}{7} & \dms{22}{40}{14}
&  4.6 $\pm$ 0.2 
& -5.3 $\pm$ 0.1  
& 0.7 $\pm$ 0.1  
& &  1.0 $\pm$ 0.1 
& -6.4 $\pm$ 0.1  
& 7.0 $\pm$ 0.3  
\\

11 &
\hmss{06}{17}{48}{7} & \dms{22}{41}{59}
&  4.0 $\pm$ 0.5 
& -6.0 $\pm$ 0.1  
& 0.9 $\pm$ 0.1  
& &  1.1 $\pm$ 0.2 
& -6.4 $\pm$ 0.1  
& 2.3 $\pm$ 0.3  
\\

12 &
\hmss{06}{17}{06}{5} & \dms{23}{01}{45}
&  5.0 $\pm$ 0.1 
& -6.0 $\pm$ 0.1  
& 1.4 $\pm$ 0.1  
& &  1.1 $\pm$ 0.1 
& -3.7 $\pm$ 0.1  
& 2.1 $\pm$ 0.1  

\enddata
\tablecomments{The velocities are in \kms.}
\tablecomments{\SICs\ 07 and 08 are positionally close and have similar systematic
velocities, and they will be labeled together as \SICs~07/08 when
necessary.}
\ifaastex
\end{deluxetable}
\else
\end{deluxetable*}
\fi


\section{Identification of Ambient Clouds Interacting with
  \icfourfourthree}
\label{sec:nature-clouds-inter}

The shocked molecular gas in \icfourfourthree\ is mostly found within
Shell A. No shocked molecular gas has
been identified outside Shell A \citep{1992ApJ...400..203D}, even
with  our new wide-field, fully-sampled mapping observations.
The signature of the shocked molecular gas is prominent along
the southern boundary of Shell A \citep[e.g.,][]{1988MNRAS.231..617B}.
On the other hand, the nature of the ambient molecular clouds that are
associated with shocked molecular gas in \icfourfourthree\ has not
been clearly understood. Our observations reveal the existence of
three different cloud complexes (\minusthreecloud, \SICs\ and
\plussixcloud) toward \icfourfourthree.  Here we investigate possible
association of these cloud complexes with \icfourfourthree.

\subsection{\SICs}
\label{sec:isolated-clumps}

\begin{figure}
\epsscale{1.}
\plotone{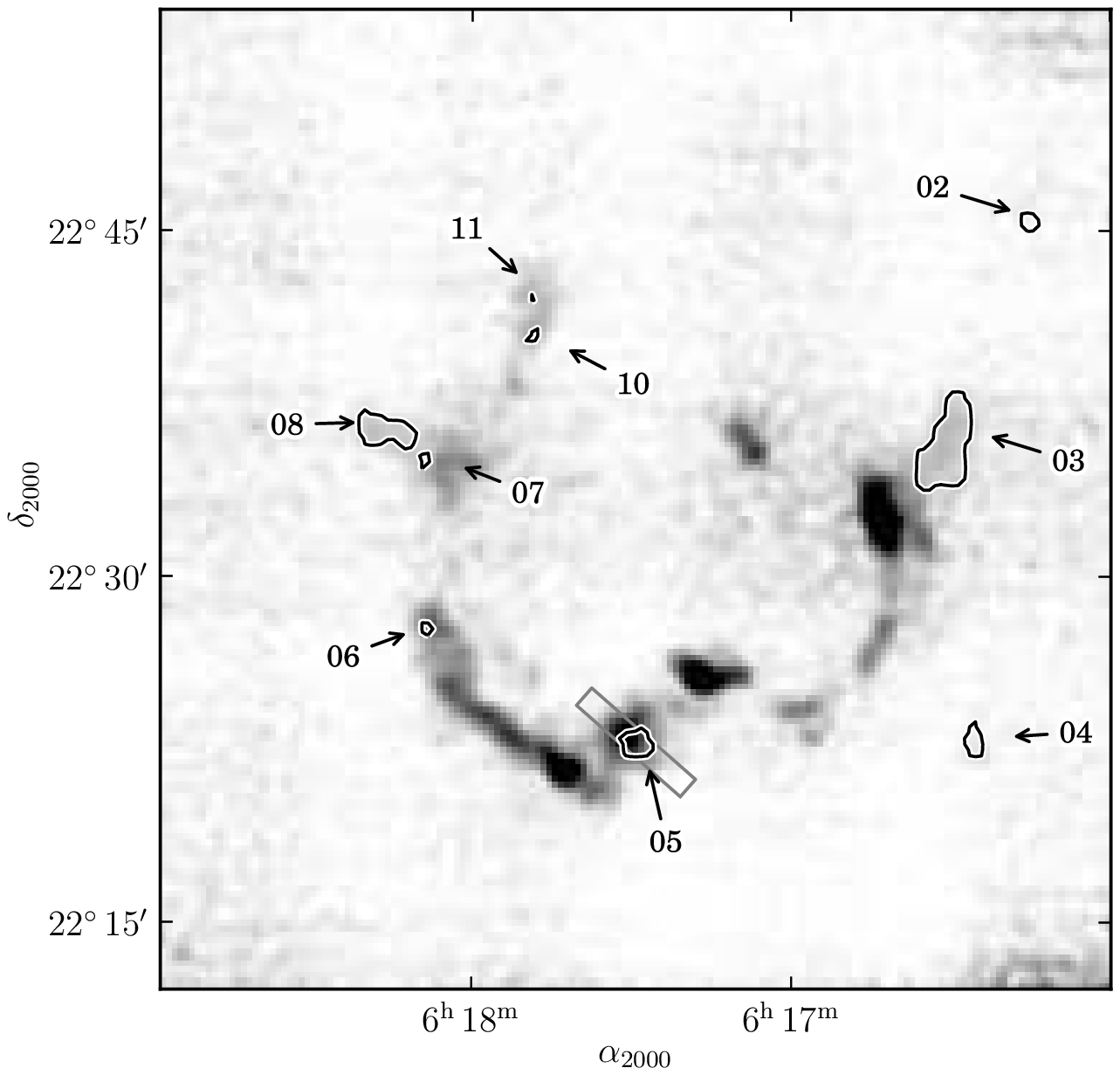}
\epsscale{1.}
\caption{ Locations of the \SICs\ overlaid over the average
  temperature map of \hcop.
The  gray rectangle is the extraction area for the PV map in
Fig.~\ref{fig:sic5-pv}.
  \label{fig:HCO+-SC}}
\end{figure}

\begin{figure}
\epsscale{0.6}
\plotone{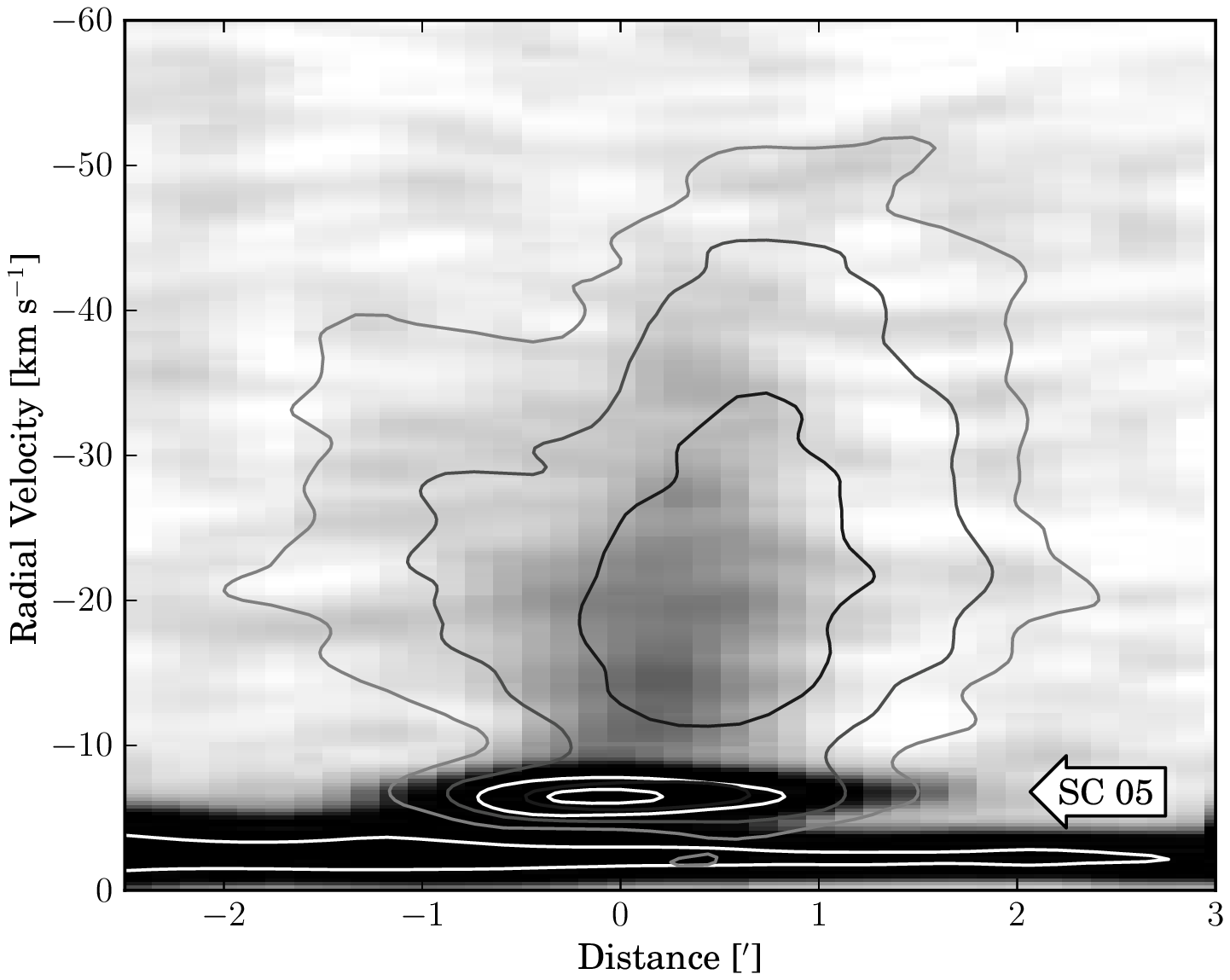}
\caption{PV map of \twelveco\ emission
  around \SIC~05 extracted from the slit in
  Fig.~\ref{fig:HCO+-SC}.
  The distance is measured from the location of \SIC~05 and the value
  increases with increasing right ascension.
  The contours represent a PV map of \hcop\ extracted from the same slit.
  The location of \SIC~05 in the PV map (distance $\sim$ 0\arcmin,
  $\vlsr\sim-6$ \kms) is marked.
  \label{fig:sic5-pv}}
\end{figure}

The SCs around \icfourfourthree\ are identified for the
first time in our observations.
We find that their spatial distribution is closely related to the
distribution of shocked gas in \icfourfourthree.  In
Fig.~\ref{fig:HCO+-SC}, we overlay the locations of the \SICs\ on top
of the distribution of the shocked molecular gas.  The positions of
\SICs\ 05, 06, 10, and 11 are coincident with those of the
shocked molecular clumps.
Also, the southeastern tip of \SIC~03 is coincident with Clump~G, and
the southwestern tip of \SICs\ 07/08 with Clump~E (we note that these
\SICs\ are spatially more extended than those regions defined by our
selection criteria).  While some of the positional coincidence could
be by chance, the fact that a large fraction of the
\SICs\ shows their projected positions on the sky being coincident
with those of shocked clumps suggests that they are more likely
physically associated.

Evidence of the physical association of the \SIC's with the shocked
molecular gas is best seen in \SIC~05.
Fig.~\ref{fig:sic5-pv} shows PV maps of
\twelveco\ and \hcop\ along an NE--SW direction and the location of this
PV cut is indicated in
Fig.~\ref{fig:HCO+-SC}.
The spatial extent of the shocked gas is nearly identical to that of
\SIC~05, unlike the emission from the more spatially extensive
\minusthreecloud.
This coincidence strongly suggests that \SIC~05 is physically related
to the shocked gas.

We further find that the
locations of the three OH 1720 MHz maser sources currently known in
\icfourfourthree\ are spatially close to the \SICs\ or even coincident
(Fig.~\ref{fig:clumps-zoom-up}).  Furthermore, the systematic
velocity of the maser
source in \SIC~05 \citep[$-6.1 \pm 0.1\ \kms$,][]{2006ApJ...652.1288H}
is coincident with the systematic velocity of \SIC~05
itself.  The velocity of the maser source near \SIC~06
\citep[$-6.9 \pm 0.1\ \kms$,][]{2006ApJ...652.1288H}
is also in agreement
with the velocity of \SIC\ 06.  The velocity of the
source near \SIC~03 (the source that is coincident with
Clump G) is $-4.55\ \kms$ \citep{2006ApJ...652.1288H}. While this
is slightly different from the peak velocity of \SIC~03, it is still
within the velocity wing of \SIC\ 03. We consider that this
maser source could be associated with \SIC~03.

Recently, \citet{2010ApJ...724.1357I} searched for H$_3^+$ absorption
features in six sight lines towards \icfourfourthree. From the
absorption features seen in two sight lines, high ionization rates
are inferred which are attributed to the increased cosmic ray
ionization rate near \icfourfourthree. The two sight lines with high
ionization rates are indeed located
close to the \SICs\ (see Fig.~\ref{fig:clumps-zoom-up}). The
centroid velocities of the absorption features are between $-8$ and
$-6\ \kms$, being consistent with the velocity range of the \SICs.

Therefore, we found that a good fraction of the \SICs\ show
direct and/or indirect association with the shocked molecular clumps.
We propose that the \SICs\ are at the same distance as \icfourfourthree,
and that some of them are in direct interaction with \icfourfourthree.

\subsection{\plussixcloud}
\label{sec:diffuse-gas-positive}

\begin{figure}
\epsscale{1.}
\plotone{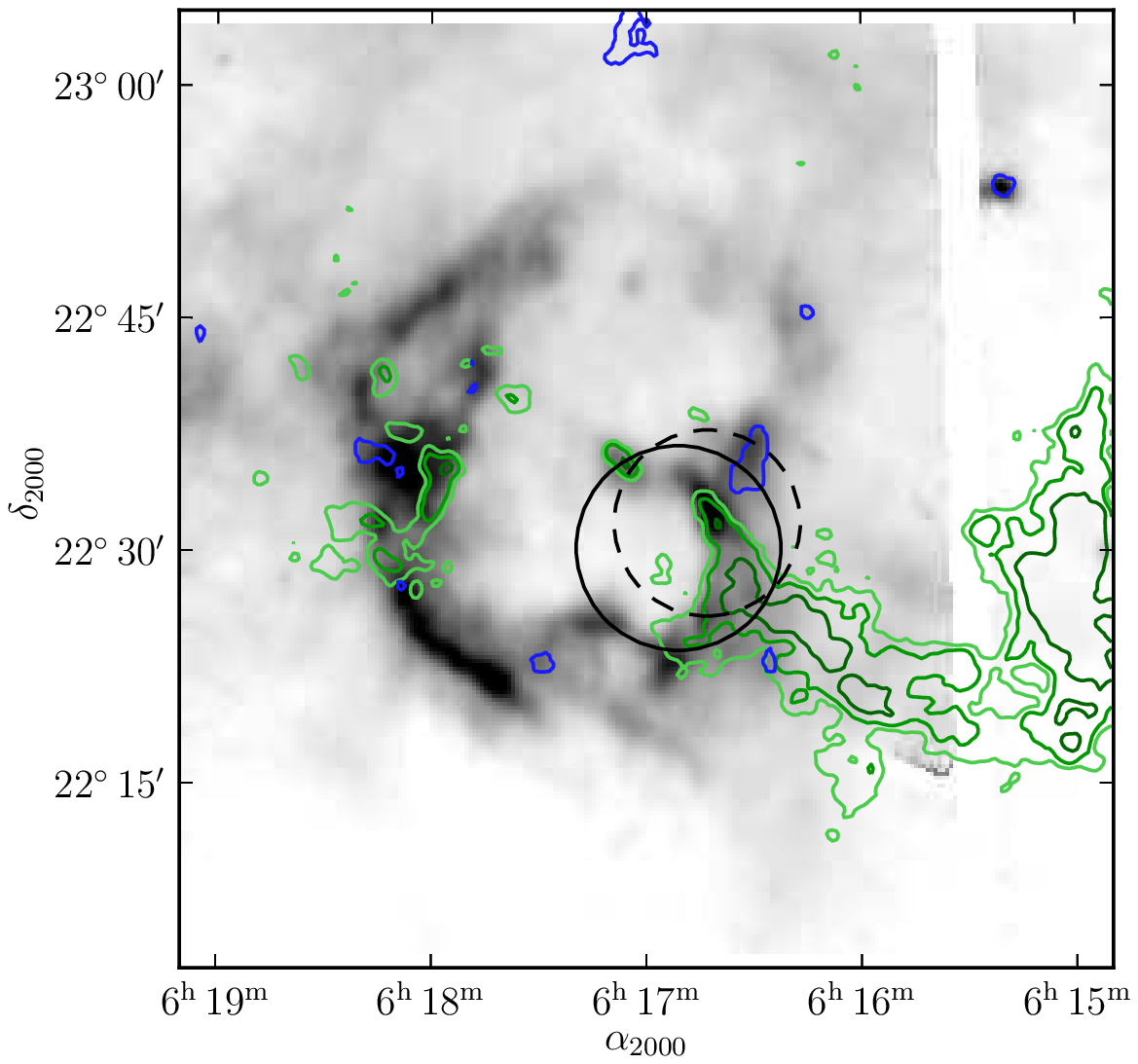}
\epsscale{1.}
\caption{
  Far IR 90 $\mu$m image taken with the \emph{AKARI}
  satellite shown in gray scale.
  The
  green contours show the distribution of \plussixcloud\ (the gray scale in
  Fig.~\ref{fig:co-positive-inter}). The blue contours show locations of
  \SICs.
  The solid and dashed circles represent the location of
  $\gamma$-ray sources detected by MAGIC and VERITAS, respectively.
  \label{fig:co-akari}
}
\end{figure}

The detailed morphology of \plussixcloud\ is also newly revealed in
our observations.  The overall morphology of the clouds, which is
shown in Fig.~\ref{fig:co-positive-inter}, is intriguing.  As
has been previously described, the clouds can be divided into two
parts. The brighter western part has a prominent filamentary
morphology. It extends from west to east. Its eastern boundary is
sharp and we see a filament of shocked gas along this boundary
(clump G and W ridge).
In Fig.~\ref{fig:co-akari}, we compare the morphology of
\plussixcloud\ to that of
90 $\mu$m far infrared emission from
\icfourfourthree. The far infrared emission, taken by \emph{AKARI}
telescope, is presumably from the shock heated dust and defines the outer
boundary of Shell A. Interestingly, the \plussixcloud\ (both the
western and the eastern parts) are only visible along and/or outside
Shell A.
Therefore, there is a good morphological coincidence between the
\plussixcloud\ and the shock tracers.

On the other hand, the systematic velocity of the
shocked gas in Clump G and the W ridge is around
$-5\ \kms$, and thus is considerable offset from that
of the \plussixcloud.  The symmetric line profile
of Clump G suggests that the shock direction is mostly
perpendicular to our line-of-sight, thus shock geometry
is unlikely to explain the 10 \kms\ difference in radial
velocity.
Can these kinematically distinct clouds be adjacent in space?
\icfourfourthree\ is likely a core-collapse SNR located in the Gem OB1
association \citep{1986A&A...164..193B},
and there could be clouds with different velocities
due to the complicated kinematics within this OB
association.
However, the difference of $10$ \kms\ seems to require rather
extraordinary condition.  Therefore, while the morphology of
the \plussixcloud\ is rather suggestive of its association with
\icfourfourthree, there is no kinematic evidence that these \plussixcloud\
are responsible for any of the shocked molecular gas. We do not
consider that the physical
association of the \plussixcloud\ with \icfourfourthree\ is conclusive.

\subsection{\minusthreecloud}

The \minusthreecloud\ show elongated morphology in the NW--SE direction,
along the interface between Shells A and B.
There is supporting evidence that most of these clouds
are located on the front side of the remnant \citep[][and references
therein]{2006ApJ...649..258T}, and the self-absorption features in
\twelveco\ and \hcop\ \citep{1993A&A...279..541V} are
consistent with this conclusion.
Thus, any interaction of \icfourfourthree\ with these
clouds would likely produce blueshifted gas; however, we see
both redshifted and blueshifted shocked gas.
Also, we see little correlation between the locations of the shocked gas
and the spatial distribution of the \minusthreecloud.  While the
velocity gradient within the \minusthreecloud\
might be due to some external
perturbation,
we could not find any hint
 that this is due to the interaction with \icfourfourthree\ or its
progenitor star.
An OH 1720 MHz maser source associated with Clump G
has a systematic velocity of $-4.55$ \kms\
\citep{2006ApJ...652.1288H}, comparable to the velocity of
\minusthreecloud. However, the location and the velocity of the maser
source
are also consistent with those of \SIC~03,
so the association of this maser source with the \minusthreecloud\ is
considered ambiguous.
We find no strong evidence for a spatial or kinematic connection
between the shocked gas and the \minusthreecloud, thus we conclude that
the remnant is not currently interacting with the \minusthreecloud.

\section{Discussion}
\label{sec:discuss}
\subsection{Origin of the Small Clouds \& Environment of \icfourfourthree}
\label{sec:envir-ic443}

\begin{figure*}
\epsscale{0.8}
\centering{
\includegraphics[scale=0.8]{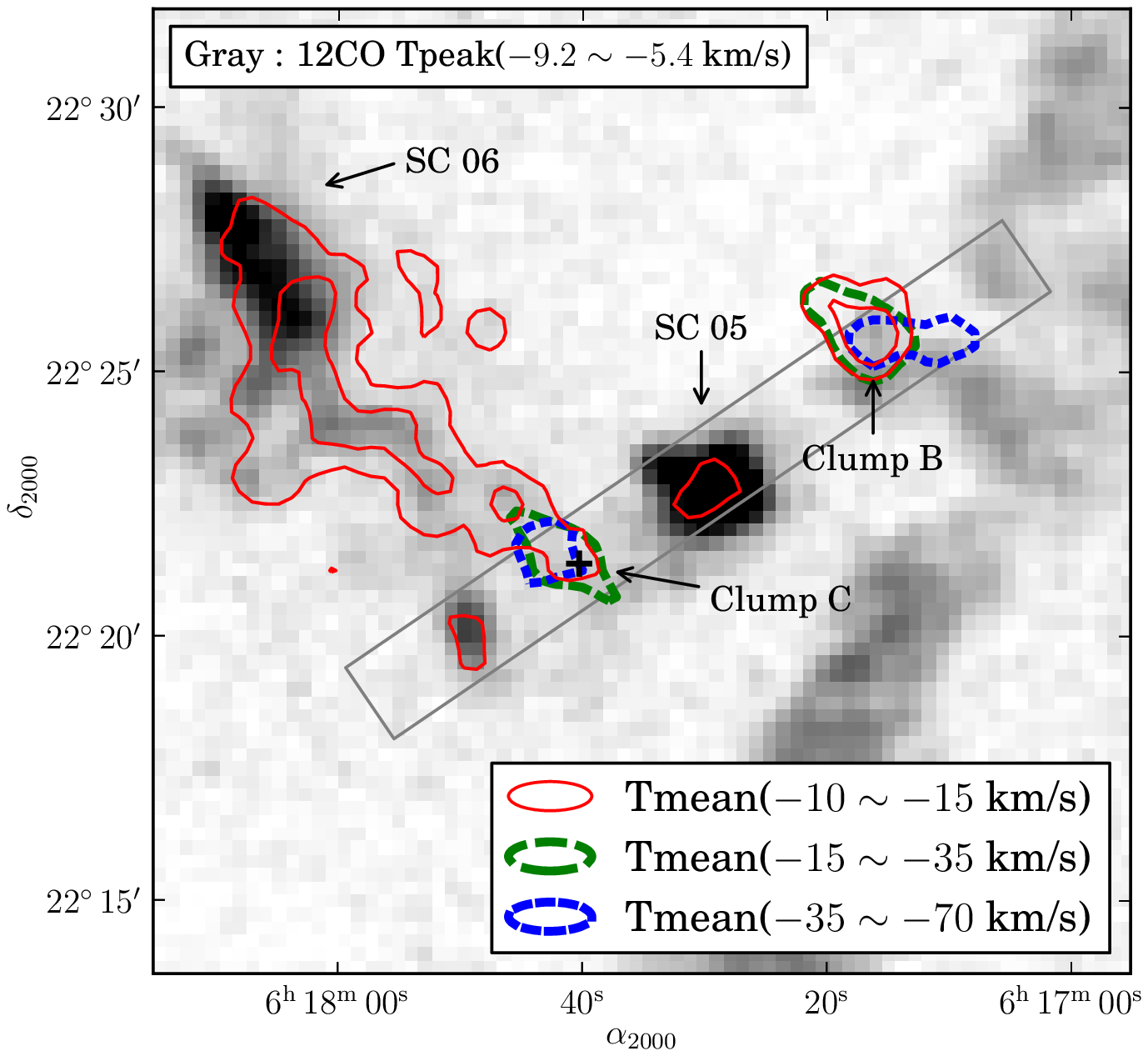}
}
\caption{ Gray scale shows a peak temperature map of \twelveco\ in
  velocity  range $-0.2$  and $-5.4$ \kms,
  mostly representing the ambient gas. Red, green, and blue contours are
  integrated temperatures of \twelveco\ for velocity range of $-10$ to
  $-15$ \kms, $-15$ to $-35$ \kms, and $-35$ to $-70$ \kms, respectively.
  They show emission from shocked molecular gas. The long
  rectangles are extraction regions for the PV maps in
  Fig.~\ref{fig:se-ridge-slit-pv}.
\label{fig:se-ridge-slit-location}}
\end{figure*}

\begin{figure}
\epsscale{0.9}
\plotone{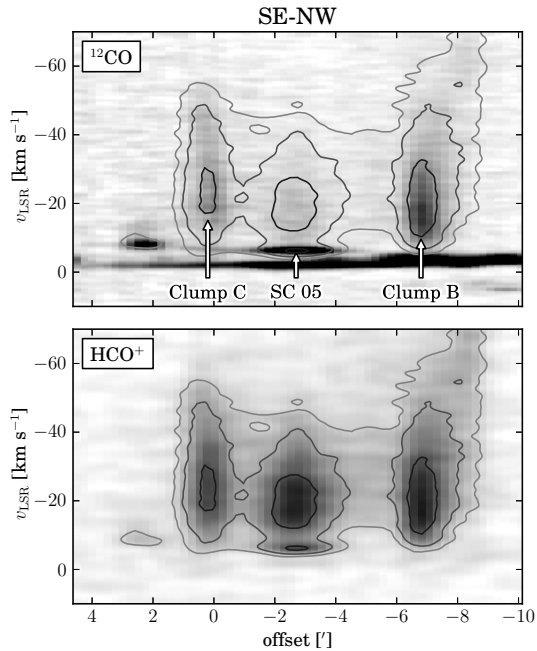}
\caption{ PV maps along the slit shown in
  Fig.~\ref{fig:se-ridge-slit-location}.
  The upper panel
  is for \twelveco, and the lower panel is for \hcop. The gray
  contours, both in upper and lower panels, are those of \hcop.  The
  offsets are distance in arcminute measured from Clump C (marked as a
  plus sign in Fig.~\ref{fig:se-ridge-slit-location}) and the offset
  increases with increasing right ascension.
  \label{fig:se-ridge-slit-pv}}
\end{figure}

\ifaastex
\begin{deluxetable}{ccc}
\else
\begin{deluxetable}{ccc}
\fi
\tablecaption{\htwo\ column densities and masses of \SICs. \label{tbl-column-density}}
\tablehead{\colhead {SC\#} & \colhead {\htwo\ Column Density} & \colhead {Mass}\\
\colhead {} & \colhead {[$\times10^{21}$ cm$^{-2}$]} & \colhead {[$\msun$]}}
\startdata

01 &
$4.3\pm0.5$ &
$12.6\pm1.6$
\\

02 &
$2.5\pm0.1$ &
$\phn3.5\pm0.1$
\\

03 &
$4.0\pm0.1$ &
$57.7\pm0.9$
\\

04 &
$3.6\pm0.1$&
$\phn7.0\pm0.1$
\\

05 &
$4.1\pm0.1$&
$13.2\pm0.2$
\\

06 &
$3.0\pm0.2$&
$\phn2.0\pm0.1$
\\

07 &
$4.2\pm0.4$&
$\phn2.7\pm0.3$
\\

08 &
$7.4\pm0.1$&
$85.8\pm0.7$
\\

09 &
$1.2\pm0.2$&
$\phn1.2\pm0.2$
\\

10 &
$1.5\pm0.1$&
$\phn0.8\pm0.1$
\\

11 &
$1.7\pm0.3$&
$\phn0.4\pm0.1$
\\

12 &
$3.2\pm0.5$&
$36.9\pm0.7$

\enddata
\tablecomments{
\htwo\ column densities are estimated from the fit parameters of component 1 in Table~\ref{tbl-peak-t-clumps}. We corrected for the main beam efficiency 
and adopted the
canonical conversion factor X ($\equiv
N_{\mathrm{H}_2}/\int{T(^{12}\mathrm{CO})dv}$) of $\sim 2 \times
10^{20}$~cm$^{-2}$K$^{-1}$km$^{-1}$s \citep[e.g.,][]{2001ApJ...547..792D}.
To estimate the mass, we multiply the \htwo\ mass by a factor of 1.36 
to account for the mass of helium.
The errors in this table are formal errors from the spectral fit and do 
not include any systematic error.
}
\ifaastex
\end{deluxetable}
\else
\end{deluxetable}
\fi


The physical characteristics of the \SICs, size of $\sim 0.5$ pc
($\sim1\arcmin$ at 1.5 kpc)
and velocity dispersion of $\sim 1\ \kms$,
resemble small clumps in molecular clouds
\citep[cf.,][]{2007ARA&A..45..339B}.
We estimate the hydrogen column density of
\SICs\ using the
canonical conversion factor X ($\equiv
N_{\mathrm{H}_2}/\int{T(^{12}\mathrm{CO})dv}$) of $\sim 2 \times
10^{20}$~cm$^{-2}$K$^{-1}$km$^{-1}$s
\citep[e.g.,][]{2001ApJ...547..792D},
and list these in Table~\ref{tbl-column-density}.
If we assume a path length of $\sim0.5$~pc (1$\arcmin$, which is a
typical angular diameter of the \SICs, at a distance of
1.5 kpc), we obtain average molecular
hydrogen densities of the order of $10^3$ \cmthree\
(higher densities must
be present to produce the \hcop\ emission seen in most of the \SICs,
suggesting that the clouds could be clumped and have a small filling
factor within the beam).
The masses of the individual clouds are estimated to be of order
of $10$ \msun\ (Table~\ref{tbl-column-density}).
This is comparable to the mass of individual shocked
clumps \citep[5--40 \msun;][]{1992ApJ...400..203D},
and also comparable to that of small clumps
\citep[cf.,][]{2007ARA&A..45..339B}.

One of the key characteristics of the \SICs\ is that they lack
surrounding clouds. We propose that the \SICs\ are analogous to the
cometary globules, i.e., remnant cores of molecular clouds that
have been exposed as their lower density gas envelopes have been
removed by the prior activity of the SN progenitor star or nearby OB
stars \citep{1983A&A...117..183R}.
For \SICs\ 01 and 12, located farther
from \icfourfourthree, the removal of lower density envelopes may have
not been complete as we see some diffuse ambient gas around them.
Some of the \SICs\ are now being impacted by the SNR
and the gas is being ablated and accelerated producing the shocked
molecular clumps  (e.g., \SIC~05 and the associated shocked gas).
The interaction of small clouds with shock waves have been comprehensively
studied by numerical simulations
\citep[e.g.,][]{1994ApJ...420..213K}. Upon interaction, transmitted
shocks will propagate into the cloud core and the envelope will be
stripped away.  This nicely explains the spatio-kinematic structure of
the gas around \SIC~05. In this picture, some part of the core is
already shocked and ablated, while the remaining part is yet to be
shocked. The ambient gas of \SIC~05 itself
is unperturbed part of the core, and the high velocity gas would
represent a flow of gas that is primarily ablated from the core.

In figure~\ref{fig:se-ridge-slit-pv}, we show a PV diagram along a slit
extending NW-SE through \SIC~05 and covers Clumps B and C and \SIC~05
(see Figure~\ref{fig:se-ridge-slit-location} for the location of the
slit). It indicates that the shocked gas associated with \SIC~05 is
quite similar to
that of Clumps B and C.
These two clumps show
more negative velocity emission than that of \SIC~05,
but no \SIC-like ambient cloud is detected for them.
Thus, we suggest that
Clump B and C represent the later evolutionary stage of the shock--core
interaction and the
cores associated with Clumps B and C are already been destroyed.
On the other hand, some of the \SICs\ (e.g., \SIC~04) do
not have associated shocked gas.
A compact ambient cloud similar to the \SICs\ is also seen near the slit
offset of $\sim 2\ \arcmin$ in Figure~\ref{fig:se-ridge-slit-pv}.  Its
characteristics are similar to that of other \SICs, except that the
peak temperature is not high enough to meet our \SIC\ criteria.
We suggest that this cloud is yet to be impacted by the
SNR shock.

The absence of small \SIC-like features with Clumps B and C points
toward the rapid destruction of the core following the passage of the
shock. The clouds are likely accelerated and destroyed within several
cloud crushing time intervals, where
the cloud crushing time is defined as $t_{cc} =
\chi^{1/2}a_0/v_b$ and $\chi$, $a_0$, $v_b$ each represents the ratio
of the density of the cloud to that of the intercloud medium, the
cloud size, and the shock velocity in the intercloud medium
\citep{1994ApJ...420..213K}.  If we adopt $a_0$ of 1 pc
 and $\chi \sim 100$, we obtain
$\tau_{cc} \sim 1\times 10^5\ (v_b/100\,\kms)^{-1}$ yrs. $v_b$ is
uncertain, but recent X-ray observations \citep{2006ApJ...649..258T}
seem to suggest $v_b\sim 500\ \kms$ while other observations tend to
give lower velocities \citep[e.g., see ][]{1999ApJ...511..798C}.
Adopting $v_b\sim 500\ \kms$ yields $\tau_{cc} \sim 2\times10^4$ yrs, which is
comparable to the estimated remnant age \citep[$\sim 30,000$
yrs,][]{1999ApJ...511..798C}. Hence the shock crushing may not be an
efficient mechanism to destroy \SICs.  However, the intercloud shock
in \icfourfourthree\ could be in radiative phase
\citep{1999ApJ...511..798C} and the large postshock density of
the intercloud shock may have significantly accelerated the destruction of
the clouds.  The \SICs\ are also subject to evaporation.
The evaporation time for classical thermal conductivity is given by
$t_{\mathrm{evap}} = 3.3\times 10^{20} n_c (\frac{a_0}{1
  \mathrm{pc}})^2 T_i^{-5/2} $ \citep{1977ApJ...211..135C}, where
$n_c$ is the hydrogen number density of the cloud and $T_i$ is the
temperature of the shocked intercloud gas.  Adopting $n_c \sim 10^3\
\cmthree$ and $T_i = 4\times 10^6$ K \citep{2006ApJ...649..258T}
yields $t_{\mathrm{evap}} > 2\times 10^6$ yrs. This is much greater than
the remnant age so the evaporation process should not have a significant
effect on the evolution of the \SICs.

On the other hand, as has been discussed in
\citet{1999ApJ...511..798C}, O-type stars likely clear out
molecular clouds farther than a 15 parsec in radius. Thus the
existence of the \SICs\ suggests that the progenitor star of
\icfourfourthree\ would not have been an O-type star. It would be more
likely an early B star (initial mass of 8--12 \msun) whose
photoionizing radiation and wind power are enough to clear out
diffuse gas but not enough to clear out dense cloud cores.

\subsection{$\gamma$-ray Emission}

\icfourfourthree\ is one of the SNRs known to be associated with
$\gamma$-ray sources. Recent observations with MAGIC
\citep{2007ApJ...664L..87A} and VERITAS \citep{2009ApJ...698L.133A}
revealed a TeV source in \icfourfourthree. The emission is seen near
Clump G (Fig.~\ref{fig:co-akari}). The origin of the $\gamma$-ray
emission in SNRs is currently not fully understood. While it is
likely from the hadronic interaction of cosmic ray protons with dense
ambient medium.
One of the keys to understanding the
origin of the $\gamma$-ray emission is the characteristics of the
ambient clouds that serve as target particles of hadronic
interactions.

The location of the detected TeV $\gamma$-ray emission is close to the
location
of the shocked clump G and this is a region where all
three components of the ambient molecular clouds are seen.  It is not
clear which cloud(s) is the target of cosmic ray protons.  The fact that
the $\gamma$-ray emission is only detected in this region implies that
the target particles are particularly abundant in this area. In this
regard, \minusthreecloud\ could not be the primary target clouds. The
recent VERITAS observation suggests that the $\gamma$-ray emission is
elongated in the north-south direction
\citep{2009ApJ...698L.133A}
and
\plussixcloud\ cloud could provide major target particles.
The intensity and spectral shape of the $\gamma$-ray emission depend
on the distance and mass of the target clouds.
Recently,
\citet{2010MNRAS.408.1257T} found that the observed $\gamma$-ray spectra of
\icfourfourthree\ could be explained by two cloud components; a cloud of
$\sim350$ \msun\ close to the remnant ($\sim 4$ pc) and a more distant
($\sim 10$ pc) massive one of $\sim4000$ \msun. The closer cloud
component may correspond to the \SIC~03  or the shocked clump G
while the distant one would be
the \plussixcloud\ or the \minusthreecloud.

\section{Summary }
\label{sec:summary}

In this paper, we present fully sampled, spectroscopic imaging of
\twelveco\ and \hcop\ molecular line emission toward the supernova
remnant \icfourfourthree\ to investigate its molecular environment and
to survey the area for additional zones of shock emission.  While all
previously known shocked clumps are detected, no additional regions of
shocks are identified in our observations.  The ambient molecular gas
toward \icfourfourthree\ shows three kinematically and morphologically
distinct components.  The \minusthreecloud\ are elongated in the
NW-SE direction along the interface between Shells A and B of \icfourfourthree.
However, there are no signposts of interaction between \icfourfourthree\ and this
molecular component.
A fainter cloud at $+6\ \kms$ is distributed
along a northeast-southwest axis.  While the morphology of its western
edge is suggestive of an interaction, the gas velocities are too
displaced from those of the shocked clumps to firmly establish its
connection to the remnant.  Twelve small clouds (\SICs) with narrow line
widths and high peak temperature are spatially coincident with shocked
molecular clumps and other tracers of interaction with the remnant.
We proposed that these \SICs\ are the residual cores of the parent
molecular cloud that have been exposed due to the radiation field and
stellar winds of the progenitor of \icfourfourthree\ and nearby OB
stars.
\icfourfourthree\
is currently encountering some of these remaining cores as indicated
by the association of the \SICs\ with shocked molecular clumps. Some
shocked molecular clumps are not associated with \SICs. In these cases, the
cores have been recently destroyed by the shock interaction.  Finally,
we note that all three cloud components are found toward the sight line of
the $\gamma$--ray source.  These components likely provide the reservoir
of protons needed to account for this $\gamma$--ray source.

\acknowledgements
We thank to the anonymous referee for valuable comments.
This work is supported by grant AST-0838222 from the National Science
Foundation.


\begin{thebibliography}{32}
\expandafter\ifx\csname natexlab\endcsname\relax\def\natexlab#1{#1}\fi

\bibitem[Abdo et al.(2010)]{2010ApJ...712..459A} Abdo, A.~A., Ackermann,
M., Ajello, M., et al.\ 2010, \apj, 712, 459


\bibitem[Acciari et al.(2009)]{2009ApJ...698L.133A} Acciari, V.~A., Aliu,
E., Arlen, T., et al.\ 2009, \apjl, 698, L133


\bibitem[Albert et al.(2007)]{2007ApJ...664L..87A} Albert, J., Aliu, E.,
Anderhub, H., et al.\ 2007, \apjl, 664, L87


\bibitem[{{Asaoka} \& {Aschenbach}(1994)}]{1994A&A...284..573A}
{Asaoka}, I., \& {Aschenbach}, B. 1994, \aap, 284, 573

\bibitem[{{Bergin} \& {Tafalla}(2007)}]{2007ARA&A..45..339B}
{Bergin}, E.~A., \& {Tafalla}, M. 2007, \araa, 45, 339

\bibitem[{{Braun} \& {Strom}(1986)}]{1986A&A...164..193B}
{Braun}, R., \& {Strom}, R.~G. 1986, \aap, 164, 193

\bibitem[{{Burton} {et~al.}(1988){Burton}, {Geballe}, {Brand}, \&
  {Webster}}]{1988MNRAS.231..617B}
{Burton}, M.~G., {Geballe}, T.~R., {Brand}, P.~W.~J.~L., \& {Webster}, A.~S.
  1988, \mnras, 231, 617

\bibitem[{{Bykov} {et~al.}(2008){Bykov}, {Krassilchtchikov}, {Uvarov},
  {Bloemen}, {Bocchino}, {Dubner}, {Giacani}, \&
  {Pavlov}}]{2008ApJ...676.1050B}
{Bykov}, A.~M., {Krassilchtchikov}, A.~M., {Uvarov}, Y.~A., {Bloemen}, H.,
  {Bocchino}, F., {Dubner}, G.~M., {Giacani}, E.~B., \& {Pavlov}, G.~G. 2008,
  \apj, 676, 1050

\bibitem[{{Chevalier}(1999)}]{1999ApJ...511..798C}
{Chevalier}, R.~A. 1999, \apj, 511, 798

\bibitem[{{Cornett} {et~al.}(1977){Cornett}, {Chin}, \&
  {Knapp}}]{1977A&A....54..889C}
{Cornett}, R.~H., {Chin}, G., \& {Knapp}, G.~R. 1977, \aap, 54, 889

\bibitem[{{Cowie} \& {McKee}(1977)}]{1977ApJ...211..135C}
{Cowie}, L.~L., \& {McKee}, C.~F. 1977, \apj, 211, 135

\bibitem[{{Dame} {et~al.}(2001){Dame}, {Hartmann}, \&
  {Thaddeus}}]{2001ApJ...547..792D}
{Dame}, T.~M., {Hartmann}, D., \& {Thaddeus}, P. 2001, \apj, 547, 792

\bibitem[{{Denoyer}(1979{\natexlab{a}})}]{1979ApJ...232L.165D}
{Denoyer}, L.~K. 1979{\natexlab{a}}, \apjl, 232, L165

\bibitem[{{Denoyer}(1979{\natexlab{b}})}]{1979ApJ...228L..41D}
---. 1979{\natexlab{b}}, \apjl, 228, L41

\bibitem[{{Dickman} {et~al.}(1992){Dickman}, {Snell}, {Ziurys}, \&
  {Huang}}]{1992ApJ...400..203D}
{Dickman}, R.~L., {Snell}, R.~L., {Ziurys}, L.~M., \& {Huang}, Y. 1992, \apj,
  400, 203

\bibitem[{{Fesen} \& {Kirshner}(1980)}]{1980ApJ...242.1023F}
{Fesen}, R.~A., \& {Kirshner}, R.~P. 1980, \apj, 242, 1023

\bibitem[{{Green}(2009)}]{2009BASI...37...45G}
{Green}, D.~A. 2009, Bulletin of the Astronomical Society of India, 37, 45

\bibitem[{{Hartman} {et~al.}(1999){Hartman}, {Bertsch}, {Bloom}, {Chen},
  {Deines-Jones}, {Esposito}, {Fichtel}, {Friedlander}, {Hunter}, {McDonald},
  {Sreekumar}, {Thompson}, {Jones}, {Lin}, {Michelson}, {Nolan}, {Tompkins},
  {Kanbach}, {Mayer-Hasselwander}, {M{\"u}cke}, {Pohl}, {Reimer}, {Kniffen},
  {Schneid}, {von Montigny}, {Mukherjee}, \& {Dingus}}]{1999ApJS..123...79H}
{Hartman}, R.~C., {Bertsch}, D.~L., {Bloom}, S.~D., {Chen}, A.~W.,
  {Deines-Jones}, P., {Esposito}, J.~A., {Fichtel}, C.~E., {Friedlander},
  D.~P., {Hunter}, S.~D., {McDonald}, L.~M., {Sreekumar}, P., {Thompson},
  D.~J., {Jones}, B.~B., {Lin}, Y.~C., {Michelson}, P.~F., {Nolan}, P.~L.,
  {Tompkins}, W.~F., {Kanbach}, G., {Mayer-Hasselwander}, H.~A., {M{\"u}cke},
  A., {Pohl}, M., {Reimer}, O., {Kniffen}, D.~A., {Schneid}, E.~J., {von
  Montigny}, C., {Mukherjee}, R., \& {Dingus}, B.~L. 1999, \apjs, 123, 79

\bibitem[{{Hewitt} {et~al.}(2006){Hewitt}, {Yusef-Zadeh}, {Wardle}, {Roberts},
  \& {Kassim}}]{2006ApJ...652.1288H}
{Hewitt}, J.~W., {Yusef-Zadeh}, F., {Wardle}, M., {Roberts}, D.~A., \&
  {Kassim}, N.~E. 2006, \apj, 652, 1288

\bibitem[Indriolo et al.(2010)]{2010ApJ...724.1357I} Indriolo, N., Blake,
G.~A., Goto, M., et al.\ 2010, \apj, 724, 1357

\bibitem[{{Jiang} {et~al.}(2010){Jiang}, {Chen}, {Wang}, {Su}, {Zhou},
  {Safi-Harb}, \& {DeLaney}}]{2010ApJ...712.1147J}
{Jiang}, B., {Chen}, Y., {Wang}, J., {Su}, Y., {Zhou}, X., {Safi-Harb}, S., \&
  {DeLaney}, T. 2010, \apj, 712, 1147

\bibitem[{{Klein} {et~al.}(1994){Klein}, {McKee}, \&
  {Colella}}]{1994ApJ...420..213K}
{Klein}, R.~I., {McKee}, C.~F., \& {Colella}, P. 1994, \apj, 420, 213

\bibitem[{{Lee} {et~al.}(2008){Lee}, {Koo}, {Yun}, {Stanimirovi{\'c}},
  {Heiles}, \& {Heyer}}]{ic443paper1}
{Lee}, J., {Koo}, B., {Yun}, M.~S., {Stanimirovi{\'c}}, S., {Heiles}, C., \&
  {Heyer}, M. 2008, \aj, 135, 796

\bibitem[{{Olbert} {et~al.}(2001){Olbert}, {Clearfield}, {Williams}, {Keohane},
  \& {Frail}}]{2001ApJ...554L.205O}
{Olbert}, C.~M., {Clearfield}, C.~R., {Williams}, N.~E., {Keohane}, J.~W., \&
  {Frail}, D.~A. 2001, \apjl, 554, L205

\bibitem[{{Reipurth}(1983)}]{1983A&A...117..183R}
{Reipurth}, B. 1983, \aap, 117, 183

\bibitem[{{Rho} {et~al.}(2001){Rho}, {Jarrett}, {Cutri}, \&
  {Reach}}]{2001ApJ...547..885R}
{Rho}, J., {Jarrett}, T.~H., {Cutri}, R.~M., \& {Reach}, W.~T. 2001, \apj, 547,
  885

\bibitem[{{Snell} {et~al.}(2005){Snell}, {Hollenbach}, {Howe}, {Neufeld},
  {Kaufman}, {Melnick}, {Bergin}, \& {Wang}}]{2005ApJ...620..758S}
{Snell}, R.~L., {Hollenbach}, D., {Howe}, J.~E., {Neufeld}, D.~A., {Kaufman},
  M.~J., {Melnick}, G.~J., {Bergin}, E.~A., \& {Wang}, Z. 2005, \apj, 620, 758

\bibitem[Tavani et al.(2010)]{2010ApJ...710L.151T} Tavani, M., Giuliani,
A., Chen, A.~W., et al.\ 2010, \apjl, 710, L151


\bibitem[{{Torres} {et~al.}(2010){Torres}, {Marrero}, \& {de Cea Del
  Pozo}}]{2010MNRAS.408.1257T}
{Torres}, D.~F., {Marrero}, A.~Y.~R., \& {de Cea Del Pozo}, E. 2010, \mnras,
  408, 1257

\bibitem[{{Troja} {et~al.}(2006){Troja}, {Bocchino}, \&
  {Reale}}]{2006ApJ...649..258T}
{Troja}, E., {Bocchino}, F., \& {Reale}, F. 2006, \apj, 649, 258

\bibitem[{{van Dishoeck} {et~al.}(1993){van Dishoeck}, {Jansen}, \&
  {Phillips}}]{1993A&A...279..541V}
{van Dishoeck}, E.~F., {Jansen}, D.~J., \& {Phillips}, T.~G. 1993, \aap, 279,
  541

\bibitem[{{Xu} {et~al.}(2011){Xu}, {Wang}, \& {Miller}}]{2011ApJ...727...81X}
{Xu}, J.-L., {Wang}, J.-J., \& {Miller}, M. 2011, \apj, 727, 81

\bibitem[{{Ziurys} {et~al.}(1989){Ziurys}, {Snell}, \&
  {Dickman}}]{1989ApJ...341..857Z}
{Ziurys}, L.~M., {Snell}, R.~L., \& {Dickman}, R.~L. 1989, \apj, 341, 857

\end{thebibliography}
\end{document}